%% file: STag_paper_v2.tex
\documentclass[twocolumn]{aastex631}

\usepackage{CJK}
\shorttitle{Supernova Tagging and Classification}
\shortauthors{Davison et al.}
\graphicspath{{./}{figures/}}

\usepackage{subfiles}
\usepackage{amsmath}
\usepackage{enumitem}
\usepackage[T1]{fontenc}
\usepackage{multirow}

\begin{document}

\title{STag: Supernova Tagging and Classification}

\correspondingauthor{David Parkinson}
\email{willdavison@kasi.re.kr,davidparkinson@kasi.re.kr}

\author{William Davison}
\affiliation{Korea Astronomy and Space Science Institute, 776, Daedeokdae-ro, Yuseong-gu, Daejeon 34055, Republic of Korea}
\affiliation{University of Science and Technology, 217, Gajeong-ro, Yuseong-gu, Daejeon 34113, Republic of Korea}

\author[0000-0002-7464-2351]{David Parkinson}
\affiliation{Korea Astronomy and Space Science Institute, 776, Daedeokdae-ro, Yuseong-gu, Daejeon 34055, Republic of Korea}
\affiliation{University of Science and Technology, 217, Gajeong-ro, Yuseong-gu, Daejeon 34113, Republic of Korea}

\author[0000-0002-4283-5159]{Brad~E.~Tucker}
\affiliation{Mt Stromlo Observatory, The Research School of Astronomy and Astrophysics, Australian National University, ACT 2611, Australia}
\affiliation{National Centre for the Public Awareness of Science, Australian National University, Canberra, ACT 2601, Australia}
\affiliation{The ARC Centre of Excellence for All-Sky Astrophysics in 3 Dimension (ASTRO 3D), Australia}



\begin{abstract}
Supernovae classes have been defined phenomenologically, based on spectral features and time series data, since  the specific details of the physics of the different explosions remain unrevealed. However, the number of these classes is increasing as objects with new features are observed, and the next generation of large surveys will only bring more variety to our attention. We apply the machine learning technique of multi-label classification to the spectra of supernovae. By measuring the probabilities of specific features or `tags' in the supernova spectra, we can compress the information from a specific object down to that suitable for a human or database scan, without the need to directly assign to a reductive `class'. We use logistic regression to assign tag probabilities, and then a feed-forward neural network to filter the objects into the standard set of classes, based solely on the tag probabilities. We present \texttt{STag}, a software package that can compute these tag probabilities and make spectral classifications.
\end{abstract}



\section{Introduction} \label{sec:intro}
\subfile{Introduction.tex}

\section{Data}\label{sec:data}
\subfile{Data.tex}

\section{Methodology}\label{sec:lr}
\subfile{Method.tex}

\section{Results}\label{sec:res}
\subfile{Results.tex}

\section{Discussion}\label{sec:dis}
\subfile{Discussion.tex}

\section{Conclusion}\label{sec:conc}
\subfile{Conclusion.tex}

\begin{acknowledgments}
The authors would like to thank Geraint Lewis and Tamara Davis for helpful comments on the manuscript. This approach was inspired by a presentation given by Lawrence Rudnick. WD and DP are supported by the project \begin{CJK}{UTF8}{mj}우주거대구조를 이용한 암흑우주 연구\end{CJK} (``Understanding Dark Universe Using Large Scale Structure of the Universe''), funded by the Ministry of Science. 
\end{acknowledgments}


%



\software{\texttt{Astropy} \citep{astropy:2013,astropy:2018},
          \texttt{DASH} \citep{Muthukrishna2019},
          \texttt{keras} \citep{chollet2015keras},
          \texttt{Matplotlib} \citep{Hunter:2007},
          \texttt{NumPy} \citep{harris2020array},
          \texttt{scikit-learn} \citep{scikit-learn},
          \texttt{SciPy} \citep{2020SciPy-NMeth},
          \texttt{TensorFlow} \citep{tensorflow2015-whitepaper}
          }


\bibliography{references}{}
\bibliographystyle{aasjournal}




\appendix
\section{Tables of Results}\label{app:res}
\subfile{res_table.tex}

\subfile{new_res_table.tex}

\subfile{SLSN_tab.tex}

\end{document}

%% file: Introduction.tex
Large surveys like the Dark Energy Survey \citep{Abbott2016} have contributed to the ever increasing number of supernova we have detected, and the Vera C. Rubin Observatory project (VRO) will significantly increase the amount beyond even this \citep{LSSTScienceCollaboration2009}. Such a wealth of data is especially important for cosmology, with Type Ia supernovae playing a critical role such as in the discovery that the expansion of the Universe is accelerating \citep{Riess1998,Perlmutter1999} and potentially for solving the ongoing tension surrounding the value of the Hubble constant \citep{Verde2019,DiValentino2021,Efstathiou2021}. There is therefore a necessity of being able to clearly distinguish Type Ia supernovae from the other types, as well as the more minute details of possible differences between the sub-types of Type Ia supernovae. Whilst photometric classification of supernovae (e.g. \citealt{Lochner2016,Charnock2017,Muthukrishna2019-2,Moller2020}) is often the easier route due to the ease of obtaining photometric data compared to spectroscopic data, spectroscopic classification is the preferred route due to its ability to identify the many sub-types of supernovae. Software like \textit{Supernova Identification} (\texttt{SNID}; \citealt{Blondin2007}) or \texttt{Superfit} \citep{Howell2005} do spectroscopic classification by using cross-correlation \citep{Tonry1979} or minimising $\chi^2$ respectively. However software like these are no longer broadly applicable, as the analysis is done by hand and the quantity of data has already reached a level where this is no longer feasible. Machine learning has become an attractive solution to solving this issue of quantity of data versus time spent analysing it, but as before many attempts so far have typically been photometric. 

There has been some success in using machine learning for spectroscopic classification, and one of the more well-known of these is \textit{Deep Automated Supernova and Host classifier} (\texttt{DASH}; \citealt{Muthukrishna2019}). \texttt{DASH} uses a convolutional neural network (CNN) in order to learn features associated with different types and ages of supernova, and is capable of accurately classifying spectra in the fraction of the time it would take using software like \texttt{SNID} or \texttt{Superfit}. By using supernova templates to train the CNN, \texttt{DASH} is reliant on accurate pre-existing classifications and so is intrinsically tied to the way supernovae are split into different classes. This means that software such as \texttt{DASH} would struggle to deal with unique or peculiar supernovae that were not part of the training sample; it is vital, then, that the system of classification is as robust as possible.

Supernovae have two main modes of explosion mechanism, either through the thermonuclear collapse of a white dwarf \citep{Maoz2014} or the core-collapse of a massive star \citep{Burrows2013}, though other mechanisms such as via electron capture \citep{Nomoto1984,Nomoto1987} are starting to become a very real possibility \citep{Hiramatsu2020}. Spectroscopic classification identifies common spectral features within a specific class, with the main three features being hydrogen lines (separating Type I and Type II), silicon lines (separating Type Ia from Type Ib/c), and helium lines (separating Type Ib and Type Ic) \citep{Filippenko1997,Blondin2007,Foley2009}. Photometric classification of supernovae involves analysing their light curves, with two of the sub-types of Type II supernovae based specifically on the shape after maximum light \citep{Turatto2003}. Indeed all types of supernovae have numerous sub-types defined for each one and further information on how these sub-types are defined can be found in \citet{Blondin2007,Silverman2012,Modjaz2016}.

The classification system commonly used is not infallible however, and there has been work in the past to further refine the classes themselves and the methods used to do the classification (e.g. \citealt{Liu2016,Prentice2017,Williamson2019}). Often these updates are trying to improve the consistency of what features are associated with what class; a common problem with how  sub-types are distinguished. Some are based on the presence of lines (e.g. TiII for Type Ia-91bg), some on the absence of lines (e.g. no Si for Type Ib/c), some on line widths (e.g. Type Ib-n; \citealt{Pastorello2007,Pastorello2008}), and others on a rather arbitrary `weak or no' lines (e.g. weak SiII and CaII H\&K for Type Ia-91T or weak/no He for Type Ic). Further complicating matters is that supernovae spectra can change with time (such as Type IIb supernovae being a transition from Type II to Type Ib as the hydrogen lines weaken \citealt{Blondin2007}). In addition to these, the `pec' sub-type consists of supernova spectra that do not fit the other sub-types of the over-arching type \citep{Blondin2007}, meaning that there currently exists no threshold for deciding when a new sub-type is introduced and when an object is merely `peculiar'.
    
When a new sub-type is introduced there may  be only a small handful of objects that fall into that class (for example \texttt{SNID} had just two template spectra for the Ia-csm sub-type; \citealt{Blondin2007}). It is highly likely that as more and more supernova are observed, the current method of classification would leave us with a plethora of sub-types with just a few objects in each class. It is therefore important to be certain that such classifications are actually representative of the true supernova population. With many of the new sub-types named after the first discovery, it is important to bear in mind that supernova spectra are not static with time and if the original example of a new sub-type undergoes some form of spectral evolution, the name may no longer make any sense.

\texttt{STag} was created to match the speed of an automated classifier like \texttt{DASH}, but to improve on the way that supernova are classified. Each spectrum is assigned tags based on the features present, and these tags are then used to determine the type of the supernova using softmax probabilities. This approach is commonly known as multi-label classification \citep{Read2011} and we incorporate both logistic regression and a neural network to achieve our goal. The aim was to create a classifier that is not based purely on comparison to previous examples,
and instead uses the specific features in the spectral information to classify supernovae. By no longer being tied to the way supernovae have been classified in the past, the inconsistencies in the system might be
avoided. The tags also have probabilities associated with them, allowing for a more rigorous check of what features are present should one desire. So \texttt{STag} provides not only  a final classification, but also a set of tags that describes the spectrum.

In Section \ref{sec:data} we describe the data used to train and to test the approach. Section \ref{sec:lr} describes the method used to assign tags to each spectra and how we use a neural network to classify each spectra. In Section \ref{sec:res} we present the performance of \texttt{STag} when applied to both supernova template data as well as real supernova data. Finally in Section \ref{sec:dis} we discuss the successes and limitations of this approach.

%% file: Data.tex
Whilst there exist a great many sub-types of supernova in the literature, in this work we use just five to simplify the process (for example \texttt{DASH}; \citealt{Muthukrishna2019} makes use of 17 different sub-types). These five sub-types are as follows: Ia, Ib, Ic, IIb, and II. We do not make any distinction between the different sub-types of Type II supernovae as only one of these, Type IIn, is done spectroscopically and is based on line widths rather than the presence or absence of a line \citep{Filippenko1997}; as of yet this is not something we have tested \texttt{STag} on being able to distinguish.

\subsection{Training Data}

\begin{figure}[ht!]
\plotone{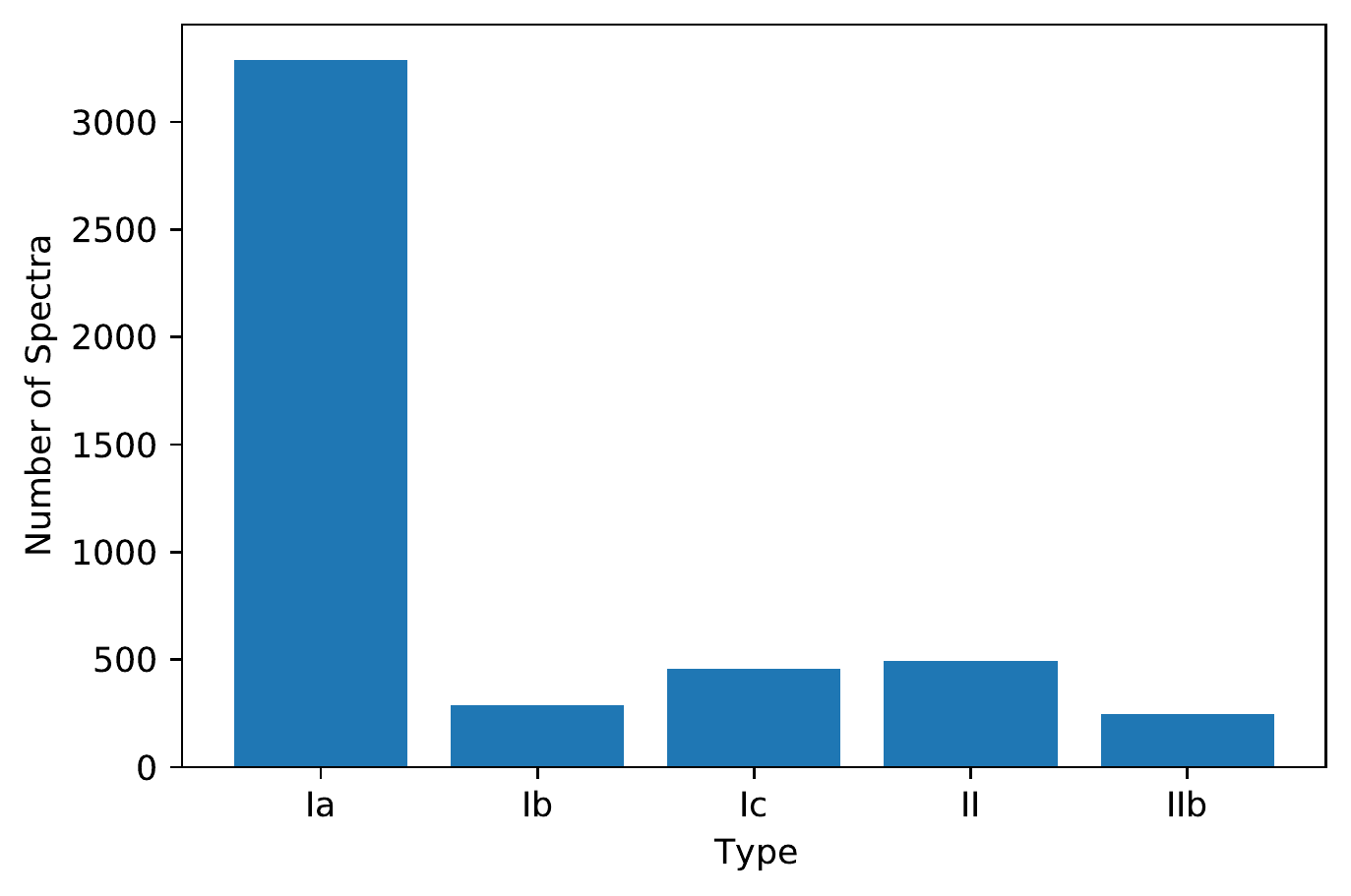}
\caption{The distribution of the supernova template spectra amongst the five classes used by \texttt{STag} before any data augmentation. We make use of a total of 4775 spectra from 509 supernovae. There are 3288 spectra from 422 Type Ia supernovae, 289 from 25 Type Ib, 459 from 37 Type Ic, 494 from 11 Type II, and 245 from 14 Type IIb. \label{fig:spec_count}}
\end{figure}

This work makes use of template spectra for training, with the templates taken from \texttt{DASH}. These were chosen as the spectra have already been labelled with their classification, streamlining the training process. The template spectra used in \texttt{DASH} consists of all types of spectra from \texttt{SNID} \citep{Blondin2007,Blondin2012}, Type Ia and Type Ic spectra from the \textit{Berkeley Supernovae Ia Program} \citep{Silverman2012}, and Type Ib and Ic spectra from a collection of supernovae by Liu \& Modjaz \citep{Liu2014,Modjaz2014,Liu2016,Modjaz2016}. The full breakdown of the number of spectra for each type is shown in Fig. \ref{fig:spec_count}, note that this includes multiple spectra from supernovae but taken at different points in time.

Since each tag of the classifier focuses on a specific feature of the spectrum, when training the logistic regression algorithm for each tag a specific sub-set of the templates was selected to improve the results. This involved selecting not only certain types/sub-types, but also hand-picking the spectra used, as not all template spectra of the same type/sub-type have the desired features present. The training sets are broken down in Table \ref{tab:pres}, with the specifics of the spectral features used as follows:

\begin{itemize}[leftmargin=0pt]
    \item[] \textbf{Type Ia -} The tags used are for the SiII line at 6150 \AA{} \citep{Silverman2012}, the CaII H\&K line at 3750 \AA{} \citep{Foley2013}, FeII at 4798-5405 \AA{} (actually a blending of multiple lines; \citealt{Silverman2012b}), and the feature(s) at 4997-5606 \AA{} (possibly part or all of two SII absorption lines; \citealt{Silverman2012b}).
    \item[] \textbf{Type Ib -} Two tags based on the HeI line at 5876 \AA{} \citep{Prentice2017}, separated depending on whether the line is absorption or emission.
    \item[] \textbf{Type Ic -} This type is in general characterised more by the lack of features than the presence of any \citep{Matheson2001}, as such no particular tag is used for this class and instead any spectrum with low probabilities for all tags or with a unique combination of tags will likely be classified as a Type Ic.
    \item[] \textbf{Type II -} The Type II supernovae are defined by the hydrogen tag, however we do not single out any specific hydrogen lines for this purpose and instead use the full spectrum.
    \item[] \textbf{Type IIb -} The tags are for the HeI line at 5876 \AA{} with a P-Cygni profile, hydrogen P-Cygni lines (most likely a H$\alpha$ line) at around 6500 \AA{}, and a double peak feature of HeI at 6678 \AA{} \citep{Turatto2003}.
\end{itemize}

\begin{deluxetable*}{cccc}
\tablecaption{Breakdown of the type of supernova and number of template spectra used to define each tag, including the wavelength(s) each tag is present at. \label{tab:pres}}
\tablehead{\colhead{Spectral Feature} & \colhead{Wavelength (\AA{})} & \colhead{Tag Present} & \colhead{Tag Not Present}}
\startdata
H & \nodata & II (494) & I (4281)\\
CaII H\&K & 3750 & Ia-norm (166) & Ic-norm (40)\\
FeII & 4798-5405 & Ia-norm (102) & Ic-norm (22)\\
SII & 4997-5606 & Ia-norm (92) & Ic-norm (40)\\
HeI (Emission) & 5876 & Ib-norm (14) & Ib-norm (78) Ic-norm (40)\\
HeI (Absorption) & \textquotedbl & Ib-norm (78) & Ib-norm (14) Ic-norm (40)\\
HeI (P Cygni) & \textquotedbl & IIb (79) & Ic-norm (40)\\
SiII & 6150 & Ia-norm (118) & Ib-norm (92) Ic-norm (40)\\
H$\alpha$ & 6500 & IIb (79) & Ia-norm (118) Ib-norm (78) Ic-norm (40)\\
HeI & 6678 & IIb (86) & Ia-norm (118) Ic-norm (22)\\
\enddata
\tablecomments{The number in brackets is the number of template spectra used of that type before data augmentation. When the $\beta$ values for each tag are determined the actual number of spectra is much larger than this. Note that the detection for the hydrogen tag is run across all wavelengths and so is not currently based on a specific hydrogen line.}
\end{deluxetable*}

\subsection{Testing Data}
During the  process of training the machine learning algorithm we set aside some fraction of the template data to validate our choices (to prevent over-fitting) and test our performance. This split between training, validation and testing data is described in Sec.~\ref{sec:res}.

For additional testing, and to allow for a consistent comparison with \texttt{DASH}, the classifier was also tested on supernova data from the Australian Dark Energy Survey (OzDES; \citealt{Lidman2020}). This sample provides a `field-test' of the approach, as each spectra is a mix of light from the supernova and host galaxy. It also has the added bonus of likely having a similar distribution of redshifts and magnitudes as would be expected from VRO. We selected 134 spectra to test, with 59 being previously classified in \citet{Muthukrishna2019} and 75 as new machine classifications.

\subsubsection{Pre-processing}
We use reduced (but unprocessed) OzDES spectra. Before any spectra can be classified, it needs to undergo a series of pre-processing steps so it is comparable to the templates used in the training data. We make use of the pre-processing methods used as part of \texttt{DASH} as outlined in \citet{Muthukrishna2019}, which are summarised as follows:

\begin{enumerate}[leftmargin=*]
    \item \textbf{Filtering:} The raw data may have cosmic rays or be particularly noisy and as such the spectrum is smoothed by way of a median filter to reduce these effects.
    
    \item \textbf{De-redshfiting:} The observed wavelengths differ from the rest frame wavelengths, so they are shifted to be in the rest frame.
    
    \item \textbf{Binning:} The wavelengths for each spectrum are binned so as to ensure that all spectra have the same number of points and the same wavelength values.
    
    \item \textbf{Continuum Removal:} A 13-point cubic spline interpolation is used to model the continuum, which the spectrum is then divided by, thus leaving only the spectral features.
    
    \item \textbf{Apodisation:} Spectra may have sharp features at their edges, so these parts of the spectrum are multiplied by a cosine filter to remove these features.
\end{enumerate}

\begin{figure}[ht!]
\plotone{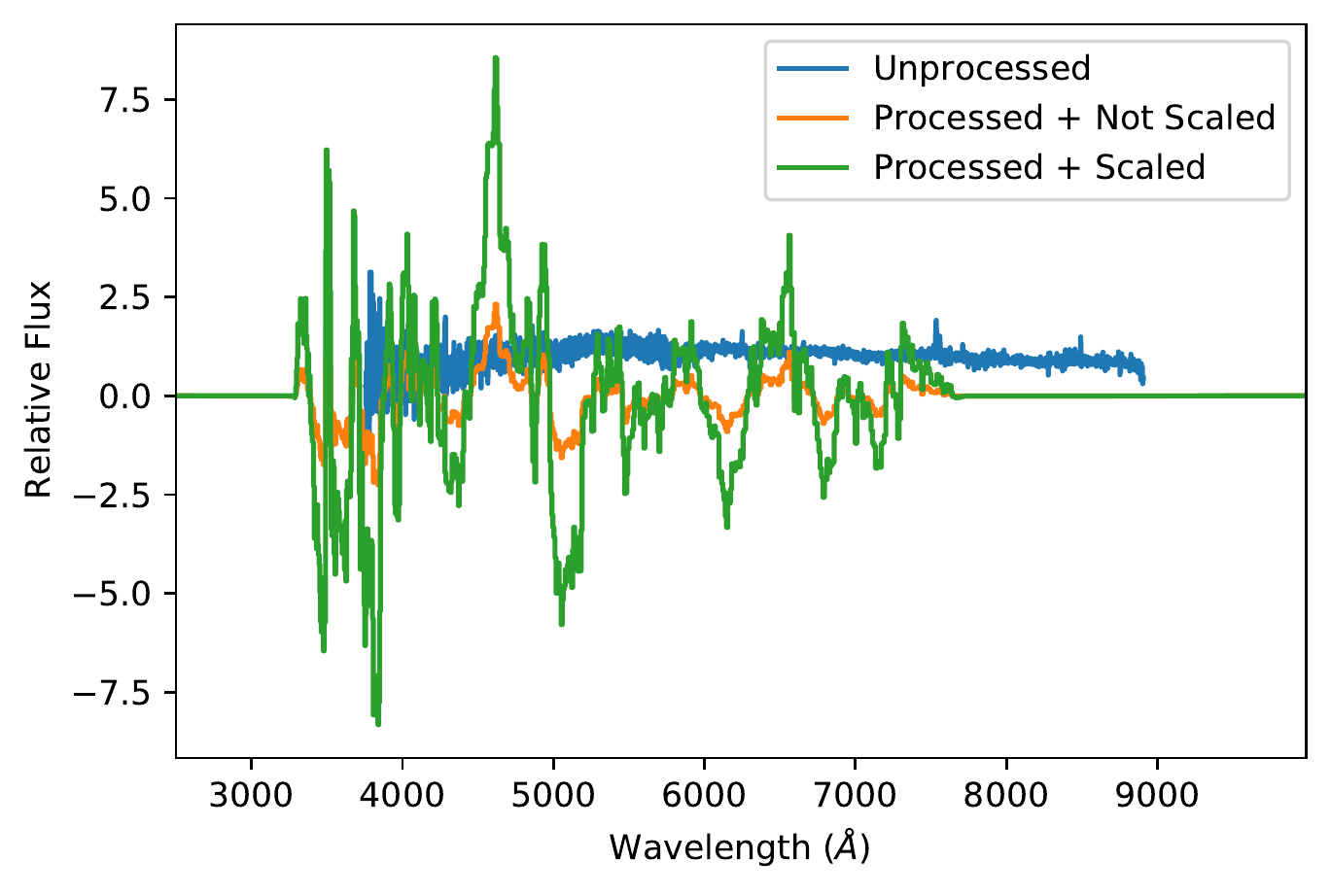}
\caption{Comparison of the spectra of \textit{DES15C2aty} taken on 15/09/17 at different stages of pre-processing. The blue line shows the spectrum of a supernova before any pre-processing steps have been applied to it. The orange line shows this data after following all of the steps as used for \texttt{DASH}, which reveals the spectral features clearly but is notably weaker in amplitude compared to the template spectra. Finally the green line shows the data after dividing through by the area under the spectrum, which scales the data to be a better match to the template spectra. \label{fig:combine_pre}}
\end{figure}

\begin{figure}[ht!]
\plotone{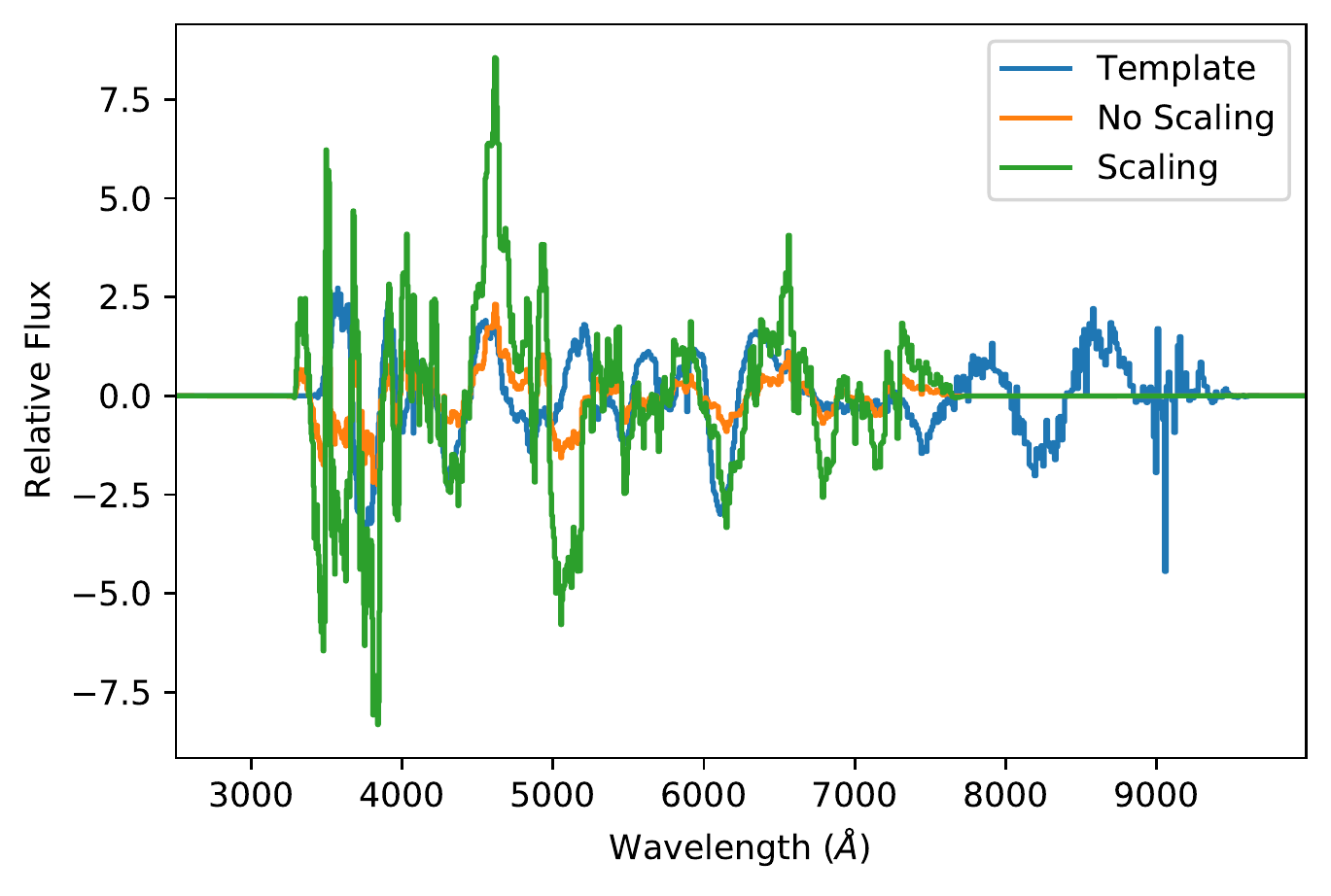}
\caption{Comparison between a template spectra, an OzDES spectra with no scaling, and an OzDES spectra with scaling. Note how the SiII absorption line at 6150 \AA{} is present in the unscaled spectra but is much smaller than the template; without the scaling this spectra would not be given a silicon tag. \label{fig:scale}}
\end{figure}

After following these steps the data spectra typically have lower flux values than the template spectra as they are from fainter objects than those that the template spectra are constructed. Since the tagging process works based on the relative strengths of the feature line to the continuum, this would lead to features being missed as they would likely be deemed too weak by the tagging process. All of the OzDES spectra are therefore enhanced by dividing through by the standard deviation before the spectrum is apodised. After apodising, the total area under the spectrum is calculated (by taking the absolute value of the fluxes so as to convert the negative flux values to positive ones) and then dividing the fluxes by this area (see Fig. \ref{fig:combine_pre} for an example of how a spectrum changes with the pre-processing steps and Fig. \ref{fig:scale} for a clearer example of how the scaling matches the features in the OzDES spectra to the template spectra). As such the OzDES spectra are scaled to match the expected parameters of the logistic regression function of each tag, reducing the possibility of introducing a false signal. We note here that the template spectra have already been pre-processed, and so are only normalised by the standard deviation of the data so that the training set has an equivalent amplitude to that of the OzDES spectra.

\subsection{Data Augmentation}
There are two problems with the template data we have available; firstly there are too few spectra overall to have enough examples to properly train the algorithm, and secondly there is an imbalance in the number of spectra for each sub-type (heavily weighted towards Type Ia). We also note that there is a reasonably small distribution in the phases of the spectra (the majority lie within $\pm30$ days of maximum light), but as the time before/after maximum light increases the strength of a supernova diminishes, and so expanding this phase range is not necessarily desirable. 

In order to solve these problems the data was artificially inflated so as to increase both the total number of spectra and also the proportion of spectra for the lesser sampled sub-types.  We augmented the data by adding Gaussian noise with mean $\mu = 0.0$ and standard deviation $\sigma^2 = 0.7$ to each flux value in the template spectra, thus essentially creating an entirely new spectrum. This value for the standard deviation was chosen so as to not significantly affect the ability of the algorithm to identify the feature(s) of interest. Each spectrum was duplicated multiple times, typically of the order 100, but in some cases up to 500 duplicates per spectrum.  

\begin{figure}[ht!]
\plotone{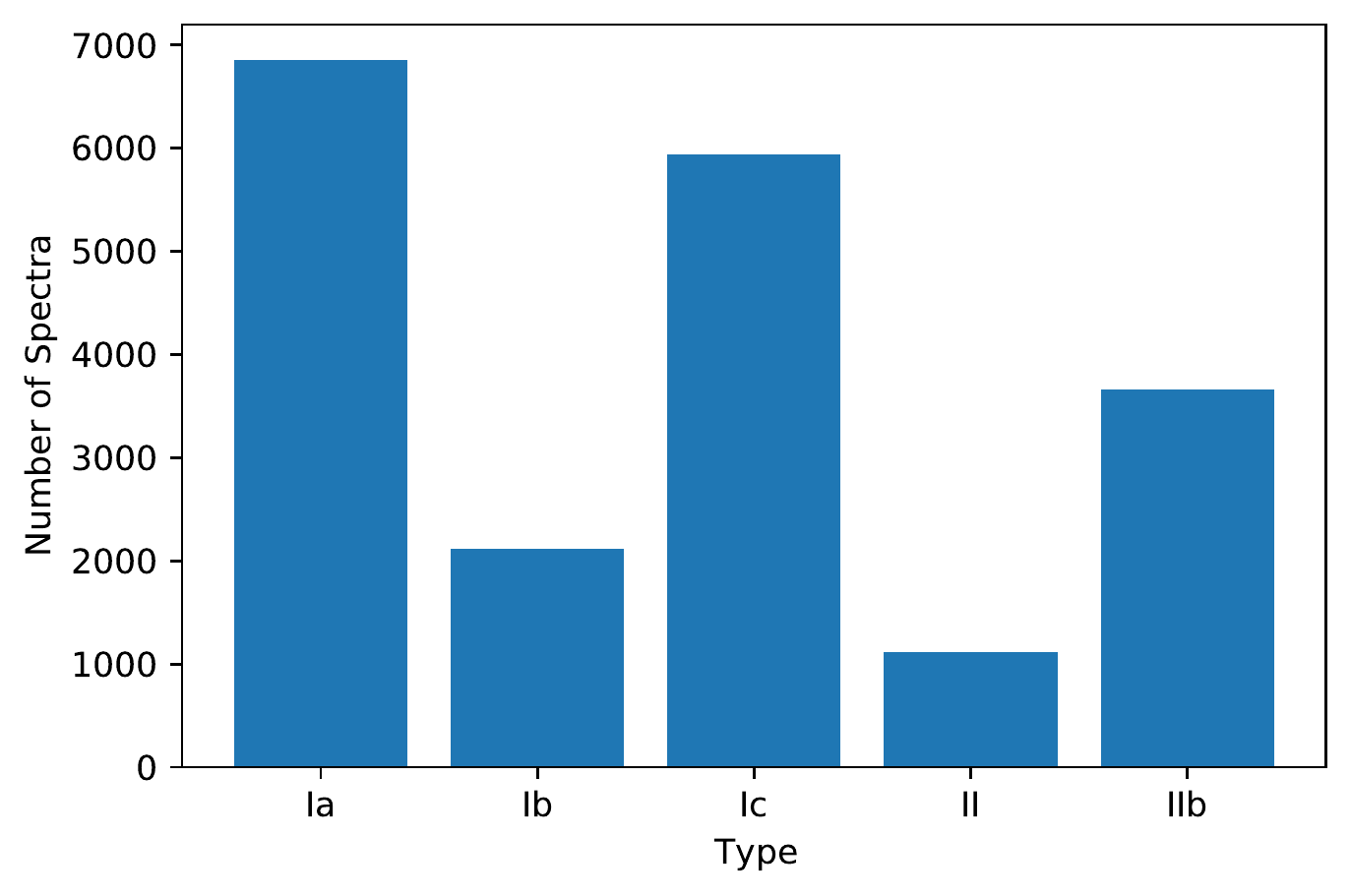}
\caption{The distribution of the supernova template spectra amongst the five classes used by \texttt{STag} after data augmentation was used for the data going into the neural network. \label{fig:spec_count_NN}}
\end{figure}

When augmenting the data for the neural network training and testing the same process is applied, though we generated a smaller amount of each type for the purposes of testing. For each tag we duplicated the template spectra 10\% of the amount done for training the tag, though as some types are repeated across tags there is still some remaining class imbalance (see Fig. \ref{fig:spec_count_NN} for the distribution of supernova used in the neural network).

%% file: Method.tex
\texttt{STag} makes use of two machine learning techniques: logistic regression \citep{logistic_regression}, used to assign tags to each spectra based on the features present, and an artificial neural network (ANN), to classify spectra based on the tags assigned to it. What results is a set of information about the spectrum, with both the tags and the class having associated probabilities. 

\subsection{Logistic Regression}
The logistic regression function, which can be used for determining the probability an object belongs to a certain class, is defined as \citep{Cramer2005}:  

\begin{equation}\label{eq:lr}
    \sigma(z) = \frac{1}{1 + e^{-z}},
\end{equation}

\noindent where \textit{z} is given by:

\begin{equation}\label{eq:z}
    z = \beta_0 + \sum^{N}_{i = 1} \beta_i x_i.
\end{equation}

\begin{figure}[ht!]
\plotone{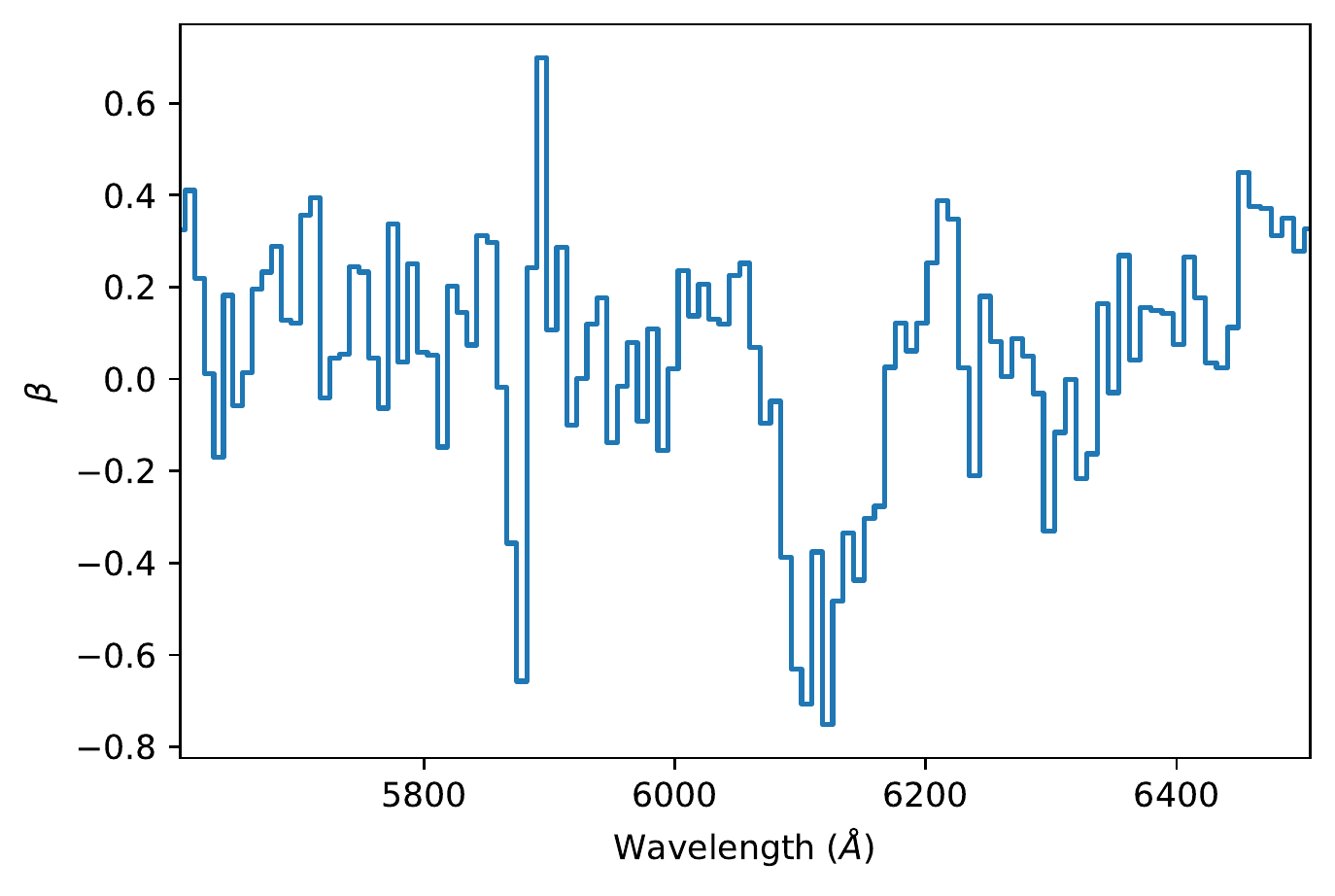}
\caption{The $\beta$ parameter values for the SiII tag, with $\beta_0$ removed as it has a much larger absolute value than the other parameters. The wide absorption peak centered at around 6150 \AA{} is the SiII line, with a possible sharp HeI 5876 \AA{} absorption line also showing. This tag does not make full use of a supernova spectra and so only has 112 parameters (including $\beta_0$). \label{fig:beta}}
\end{figure}

\noindent Here $\beta_0$ is effectively a normalisation constant, whilst the remaining $\beta_i$ values are the weights associated with values of some variable $x_i$ (in this case the flux values). We note here that the identification of features is not based on some signal-to-noise ratio, rather the $\beta$ values are representative of the actual shape of the spectral feature and so it is important to use spectra with clearly defined features in order to train a tag that is suitable; see Fig. \ref{fig:beta} for an example of the $\beta$ values for the SiII line tag.

\begin{figure}[ht!]
\plotone{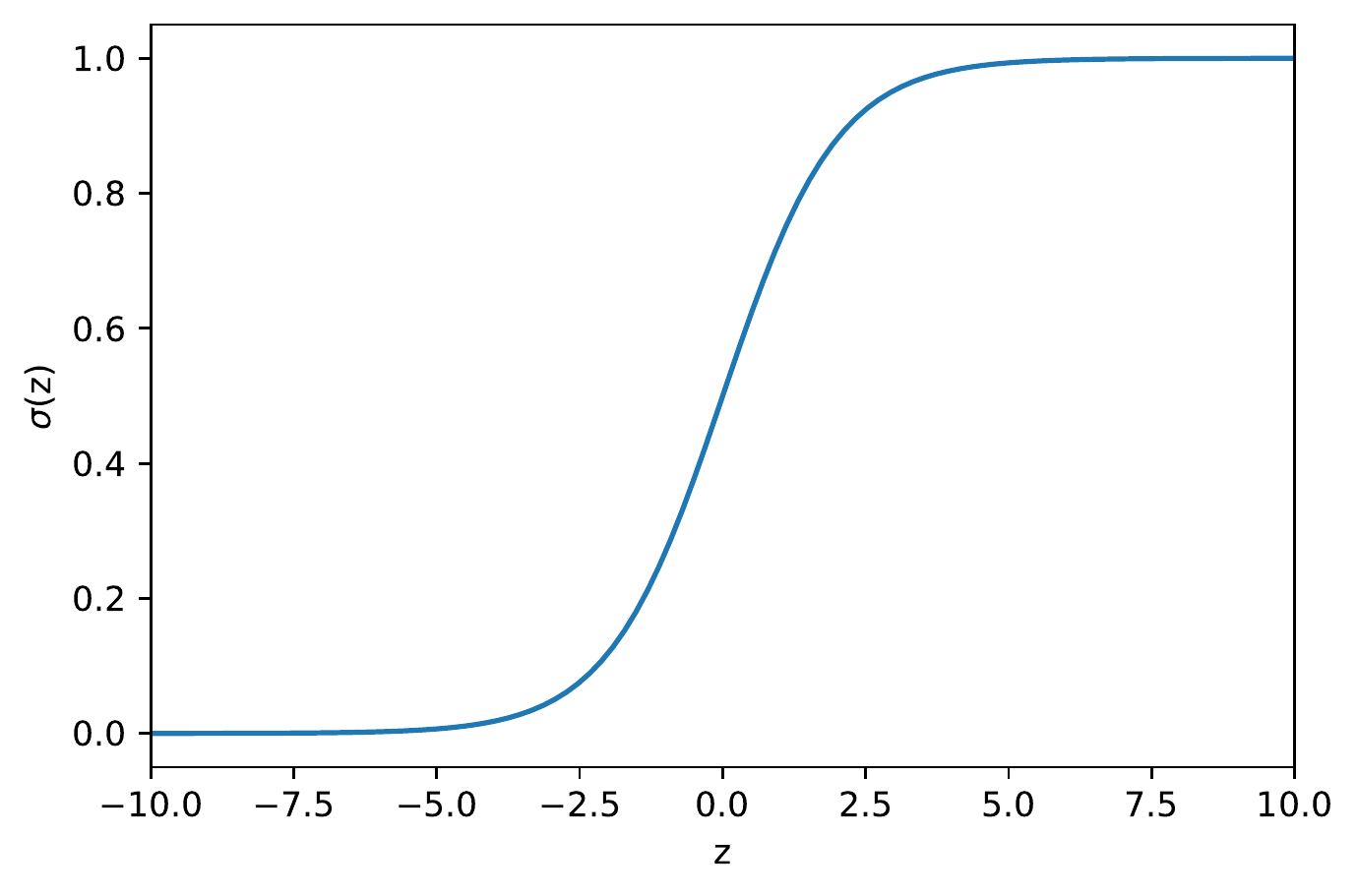}
\caption{The form of the logistic regression function, where \textit{z} is the sum of the $\beta$ parameters multiplied by the flux values and $\sigma(z)$ is the probability of the spectra having the tag. \label{fig:lr}}
\end{figure}

Logistic regression has the form as seen in Fig. \ref{fig:lr}, showing a sharp transition from 0 to 1, making it suitable for use in binary classification problems. If an object has a $\sigma(z)$ value very close to 1 it has a high probability of belonging to the class, whilst a value that is close to 0 suggests a low probability of belonging to the class. Here logistic regression is used to determine the probability that a spectrum has the feature we are trying to tag. We combine logistic regression with the cross-entropy loss function in order to optimise the weights of Eq. \ref{eq:z} for each tag. The cross-entropy loss function is given by:

\begin{equation}\label{eq:ce}
    J = - \frac{1}{N} \sum^N_{i=1} [Y_i\ln{\sigma(z)_i} + (1-Y_i)\ln{(1-\sigma(z)_i)}],
\end{equation}

\noindent where \textit{J} is the loss, \textit{N} is the number of objects to be tagged, and $Y_i$ is equal to 1 if the object has the tag or 0 if the object does not have the tag. We make use of binary relevance multi-label classification to tag for each feature, whereby each spectra has the probabilities for each tag determined independently of the other tags \citep{Read2011}. The final $\beta$ values for a tag are calculated by minimising the loss function, with a value as close to 0 as possible being the most desirable.

To optimise the loss function we employ stochastic gradient descent (SGD) as it is well suited to working with large datasets because of its reduction in running time with minimal cost to accuracy \citep{Bousquet2007,Bottou2010}. SGD differs from regular gradient descent by only using a small subset (known as a batch) of the training data to modify the gradients. This requires the training data to be fully representative, and the batch size (the number of samples in the batch) to be large enough to capture this. We found that a batch size of 2000 was sufficient to reduce the value of the loss function to an acceptable level.

\subsection{Tags}
\begin{figure*}[ht!]
\plotone{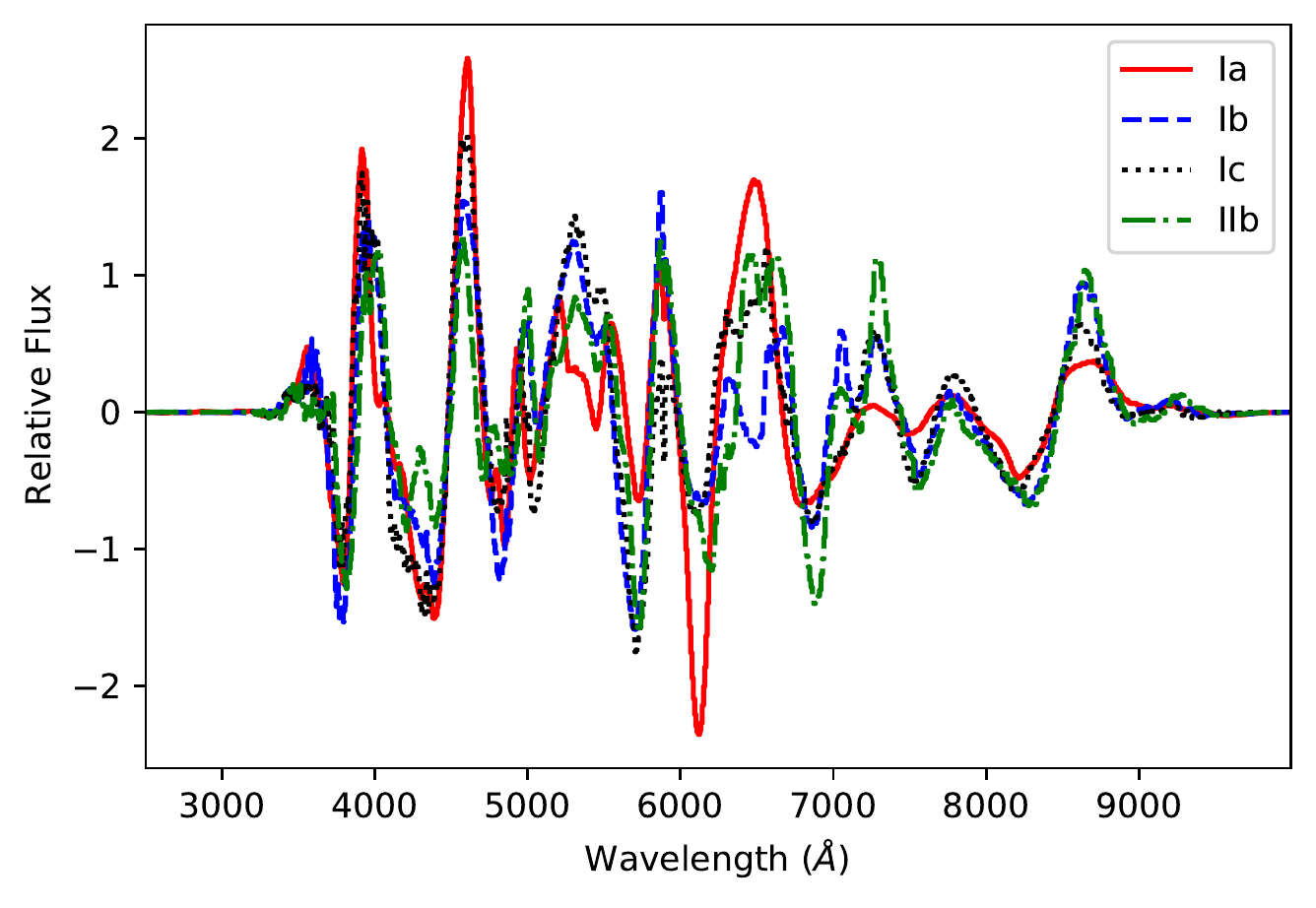}
\caption{The mean continuum-subtracted flux values of four of the supernova classes, which was used to identify features specific to a particular type. We didn't include Type II for this plot as it was deemed that the general hydrogen tag was more than sufficient to identify such cases. \label{fig:type_comp}}
\end{figure*}

We make use of many of the common features used to identify supernova as tags whilst also selecting a handful of other features unique to one of the five classes we are using; see Table \ref{tab:pres} for the full list and their associated wavelengths. Fig. \ref{fig:type_comp} shows a plot of the mean continuum-subtracted flux values of four of the five classes using template spectra, which was used to identify useful features for tagging. It is important that a feature is unique to a class and also strong enough to be detected in real (noisy) spectra.

\subsection{Artificial Neural Network}
\begin{figure*}[ht!]
    \plotone{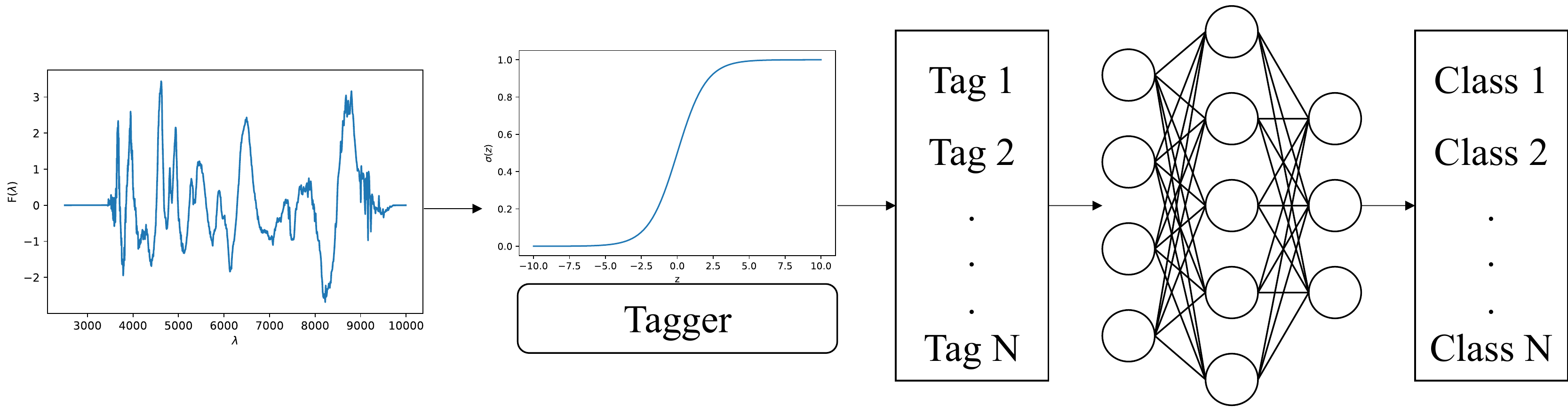}
    \caption{A simplified diagrammatic overview of the general architecture used by \texttt{STag}. The first stage is to take some pre-processed spectra and use multi-label classification by way of the logistic regression in order to assign probabilities for all of the tags. These tags are then passed to an artificial neural network which uses softmax regression in order determine the most suitable class for the supernova.\label{fig:stag}}
\end{figure*}

Differing to \texttt{DASH}, which makes use of a convolutional neural network \citep{Muthukrishna2019}, we use a simple ANN (specifically a feedforward neural network) to classify the supernova based on their tag probabilities; the full procedure is summarised in Fig. \ref{fig:stag}. We chose a a neural network specifically due to the ease of scalability in the future as and when we implement more tags and/or classes. We make use of the \texttt{TensorFlow} \citep{tensorflow2015-whitepaper} \texttt{python} implementation of \texttt{keras} \citep{chollet2015keras} to build the neural network. The neural network for this work consists of an input layer containing the same number of nodes as there are tags we are classifying on, in this case 10. We then have one hidden layer made up of 20 nodes and finally an output layer which has 5 nodes, one for each of the 5 classes we are using. The number of nodes in the hidden layer was chosen to be 20 as this gave the overall best performance before diminishing returns began. Each of the layers in this architecture is dense, which means that each node in a layer is connected to every other node in the subsequent layer. The input and hidden layers use the Rectified Linear Unit (ReLU) activation function \citep{nair2010rectified} and the output layer uses softmax, defined as:

\begin{equation}
    s(x_i) = \frac{e^{x_i}}{\sum\limits_{j=1}^N e^{x_j}}.
\end{equation}

\begin{figure}[ht!]
    \plotone{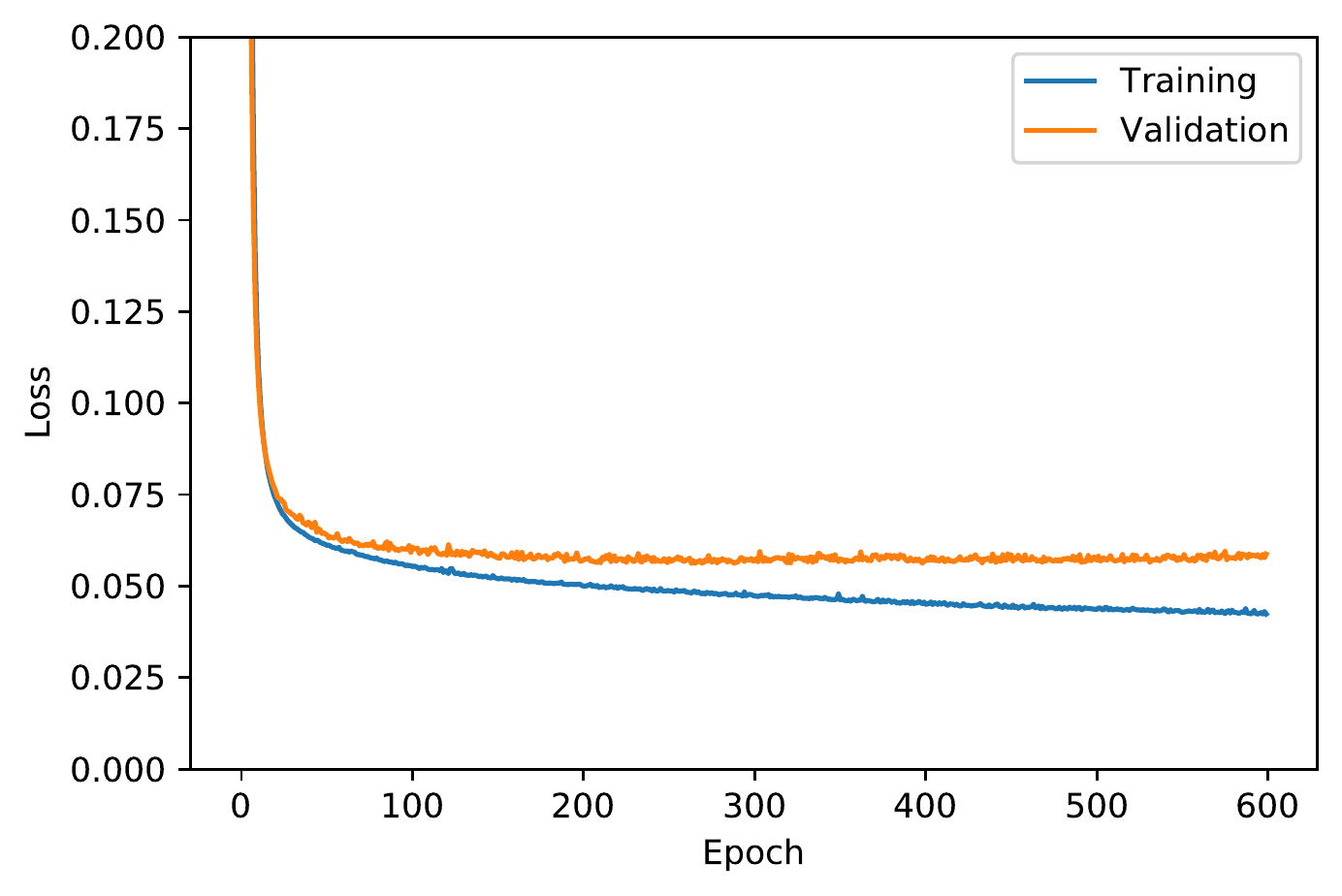}
    \caption{Comparison between the training loss and validation loss of the neural network, plotted against the epoch. Whilst the two lines start diverging earlier than the final epoch, we require a minimum of 600 epochs to achieve a 99\% success rate for classifying Type Ia supernovae. The loss difference is only small at this point, but further epochs would likely see this discrepancy continue to grow. \label{fig:loss}}
\end{figure}

\noindent Here $x$ is a vector of inputs based on which the classification occurs (in this case they are the weights of the five nodes in the output layer) and the output softmax is a vector of probabilities of length equal to the number of classes. Each probability in the softmax corresponds to the probability the object belongs to a particular class, with the highest probability being taken to be predicted class, and the sum of all probabilities equal to 1. We trained the neural network for a total of 600 epochs with a batch size of 250 and a learning rate of 0.001. We make use of the \textit{Nadam} optimiser \citep{Dozat2016} as this proved to give the best performance amongst all the optimisers we tested. We choose to stop training after 600 epochs as this is the point at which we are able to achieve a 99\% success rate for classifying Type Ia supernovae and the loss curves indicate that for more epochs the neural network will start overfitting (see Fig. \ref{fig:loss}).

In principle, the neural network learns that a certain class is associated with certain probabilities of each tag, and so is able to accurately predict what type it is. For example, members of the Ia class should have a silicon tag with a high probability whilst the other types will have low probabilities for this tag. As such,  any spectra with a high probability for this silicon tag should be heavily favoured to being predicted as a Type Ia.

%% file: Results.tex
We tested \texttt{STag} against template spectra with added noise,  spectra that were previously classified by OzDES through Astronomical Telegrams (ATels; \citealt{2015ATel.8164....1B,2015ATel.8367....1D,2015ATel.8167....1L,2015ATel.8176....1S,2015ATel.8137....1T}) and \texttt{DASH}, and finally OzDES spectra that had been classified through ATels but not in \citet{Muthukrishna2019}, with the exception of \textit{DES15E1ece} which was classified separately from these as part of the Dark Energy Survey \citep{2015ATel.8092....1D,2015ATel.8177....1G,2015ATel.7973....1K,2015ATel.8079....1S,2016ATel.8952....1C,2016ATel.8954....1C,2016ATel.8658....1G,2016ATel.8707....1K,2016ATel.8564....1P,Smith2020}.

The ATel classifications were done manually through the use of \texttt{SNID} and \texttt{Superfit}, and also made use of photometric data in addition to the spectra \citep{Muthukrishna2019}. The ATel classifications are not as detailed as the ones from \texttt{DASH}, typically lacking sub-type information and a less precise estimate of the age of the supernova. In this work \texttt{STag} does not make multiple sub-type classifications for each main type (with the exception of the Ib Type). This was an intentional decision in order to investigate the ability to classify based on spectral features; the addition of further tags would allow for the inclusion of more sub-type classifications as well. We also do not make any age predictions, though again this was an intentional decision to simplify the process before expanding it. As such we do not include the age estimates from either ATel or \texttt{DASH}. By comparing with real data we can get a good idea for how well \texttt{STag} will be able to classify supernova spectra from upcoming surveys like VRO.

\subsection{Comparison with Template Spectra}
\begin{figure}[ht!]
\plotone{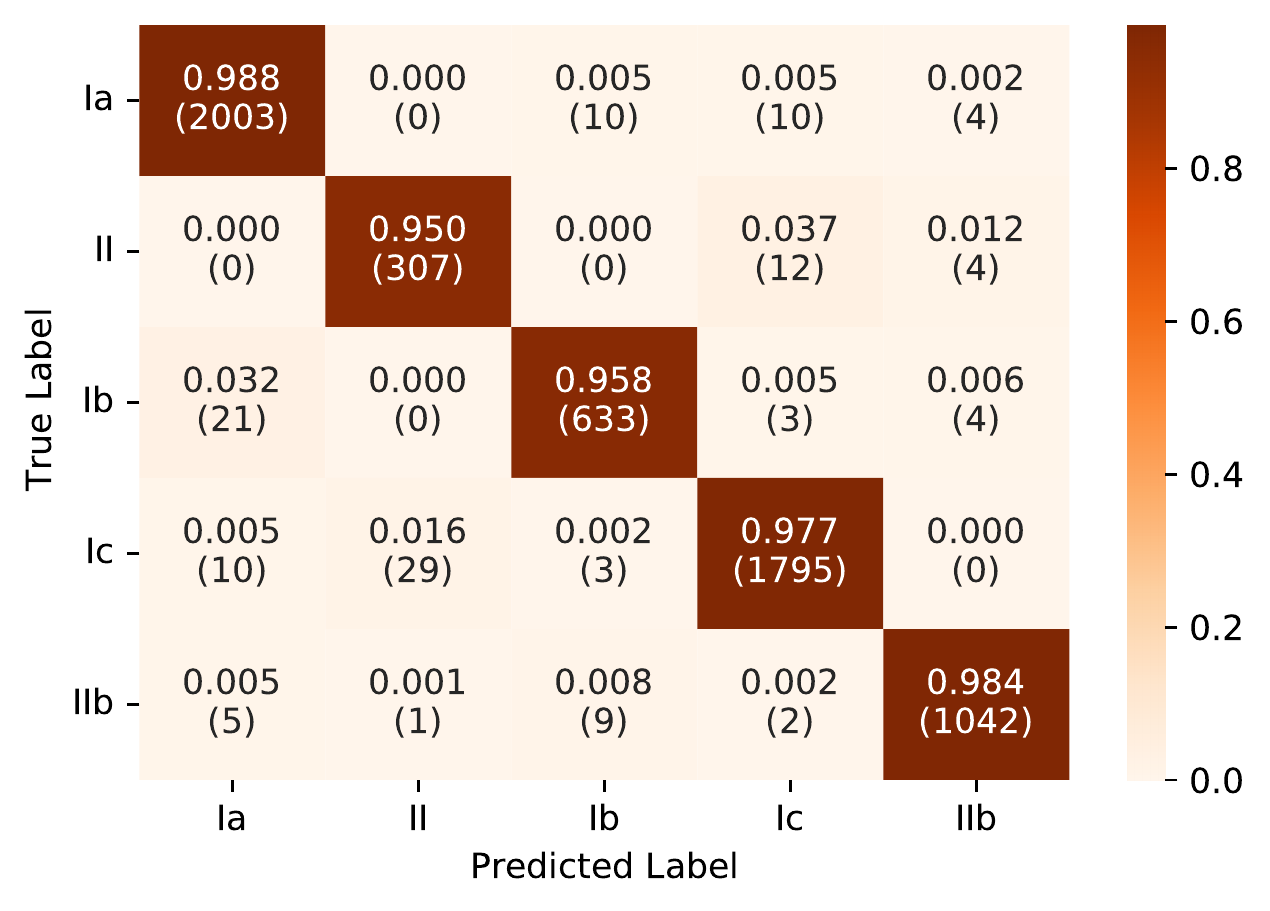}
\caption{The normalised confusion matrix comparing the predicted class made by \texttt{STag} to the `true' class of the supernova (this being the one originally assigned to the template spectra) for a total of 5907 different spectra. The decimal numbers within the squares, as well as the corresponding colours, show the fraction of predicted classes that were in agreement with the `true' class, with the number of spectra in brackets below. \label{fig:cm}}
\end{figure}

For the augmented data generated for use in the neural network as described in Section \ref{sec:data}, we used 56\% as training data, 14\% as validation data, and the remaining 30\% for testing the trained model; these values were chosen as they are reasonably standard for machine learning problems. The normalised confusion matrix of the testing data is shown in Fig. \ref{fig:cm}, which gives a visual representation of how the class predicted by the neural network compares to the real class of the original template spectra.

The trained neural network is highly accurate, capable of achieving at least a 95\% success rate for all classes, with this accuracy only expected to increase as more tags are introduced for the different classes.  

\subsection{Previously Classified Spectra}
We test the performance of \texttt{STag} against 59 OzDES spectra that have been classified through ATels and by \texttt{DASH}. The comparison between all three classifications can be seen in Table \ref{tab:comp} found in Appendix \ref{app:res}, which also includes the probabilities of any tags that are above 10\% for each spectra. Some ATel classifications are followed by a question mark to indicate that it was not a confident classification and any \texttt{DASH} classification that is followed by an asterisk means that it was unreliable. We find that we are in agreement with the ATels classifications 86\% of the time and with the \texttt{DASH} classifications 83\% of the time. We will now discuss the cases where the classifications did not match the classifications from \texttt{DASH}.

\subsubsection{DES15S2ar}
Although both the ATel (albeit not confidently) and the \texttt{DASH} classification were a Type Ia, \texttt{STag} gave a very strong Type Ic classification. Given the tags this may seem slightly strange, as it does have high probabilities for many of the tags associated with a Type Ia. However the spectra from 15/09/19 has both a low probability presence of the SiII feature (believed to be the most important tag for the Type Ia classification) and for the He-6678 feature (which is unlikely to be present at any significant level in a typical Type Ia spectra). The spectra from 15/09/21 has a very low (sub 10\%) probability for the He-6678 tag, as well as a very low probability of the SiII line, and as such this combination is highly unlikely for a Type Ia spectra. 

\subsubsection{DES15C3bj}
Our classification of Type II agrees with the (admittedly not confident) ATel classification of this supernova, and may not necessarily disagree with the one from \texttt{DASH} either. \texttt{DASH} found this supernovae to be a Type Ic-broad, a classification considered unreliable as \texttt{DASH} tends to classify host-contaminated spectra as Ic-broad \citep{Muthukrishna2019}. The 100\% hydrogen tag for the two spectra means that we are confident that \texttt{STag} is making the correct classification.

\subsubsection{DES15S1by}
Both \texttt{STag} and the ATel classification give this as a Type II, whereas \texttt{DASH} classifies this as a Type Ic.  However the \texttt{DASH} classification of this supernova was considered to be unreliable, and it also has the same issue with being classed as Ic-broad much like \textit{DES15S1bj}. The 100\% hydrogen tag (as well as very few other tags) for both spectra is the clear reason for the Type II classification by \texttt{STag}. This agreement with the ATel classification means that it is a classification we are confident with.

\subsubsection{DES15X3dyu}
Whilst typically the hydrogen tag would cause a Type II classification, the lack of the H$\alpha$ tag combined with the strong SiII tag makes for a unique combination that is unlike other Type II supernovae and so is likely why we get a Type Ic classification. This hydrogen tag is no longer present in the second spectrum, potentially highlighting this is a supernova type with an evolving spectrum. For example supernovae of the Ia-csm sub-type have hydrogen lines in their spectra due to circumstellar material \citep{Silverman2013}, which could be a possible source of the lines in this spectra and could weaken over time.

\subsection{DES15C3edd}
Both the ATel and \texttt{DASH} classifications were that of a Type Ia, however for the earlier spectrum \texttt{STag} classifies this supernova as a type Ib, though not particularly confidently. For the later spectra we get a 100\% Type Ia classification, and this discrepancy can be accounted for by the lack of the SiII tag and low probability of the He-6678 tag in the earlier spectra which cause what we believe to be a misclassification by \texttt{STag}.

\subsubsection{DES15C3lvt}
This supernova seemingly does not have a very reliable classification from any of the three sources used in this work. The ATel classification is an unreliable Type Ia, the one from \texttt{DASH} is an unreliable Type Ic, and for the first spectra we get a fairly unreliable (64.8\%) Type II classification from \texttt{STag}. The one from \texttt{STag} is particularly puzzling as there is no hydrogen tag, so whilst it seems there is no good fit to any of the classes, it is not clear why this wasn't classified as a Type Ic. The second spectrum has a much more confident Type Ib classification from \texttt{STag}, though given the unreliability of the previous attempts further investigation would be desirable before committing to this assigned class.

\subsection{Newly Machine Classified Spectra}
We also tested \texttt{STag} on 75 OzDES spectra that had not been classified in \citet{Muthukrishna2019} and only some have ATel classifications. As such the redshifts for the cases that had no prior classification were calculated using \texttt{\textsc{Marz}} \citep{Hinton2016} and then all spectra were classified with \texttt{DASH} and \texttt{STag}; the results of these classifications and the associated tags can be found in Table \ref{tab:comp2} of Appendix \ref{app:res}. As before, \texttt{DASH} classifications followed by an asterisk indicates the classification was unreliable. We find that we are in agreement with the classification from \texttt{DASH} only 39\% of the time, which whilst at first glance is a low level of agreement we note that a significant portion of the \texttt{DASH} classifications have been flagged as unreliable. Although many of the classifications are not in agreement, we believe the \texttt{STag} classifications to be the more believable in such cases, because so many from \texttt{DASH} are flagged as unreliable. Whilst we will not cover all the cases that show disagreement, we will instead explain some of the more common causes for different classifications.

There are some instances where \texttt{DASH} predicts a Type IIb and \texttt{STag} predicts a Type II, which in the latter case is caused by a strong hydrogen tag. Since Type IIb supernovae initially show hydrogen in their spectra, which weakens over time, it is possible that in these cases  the lack of age information in \texttt{STag} is the limiting factor, although it also likely highlights the limitation of having a transitioning class. We also note that the Type Ic class for \texttt{STag} is the class an object goes if there is no good fit, so if there are other important spectral features we don't have tags for, this could explain cases where \texttt{STag} predicts Type Ic and \texttt{DASH} predicts a different class.

We also include two supernova spectra that have previously been classified as SLSN-I, a type of superluminous supernova that show little to no sign of hydrogen \citep{Gal-Yam2012}. Since neither \texttt{DASH} nor \texttt{STag} were trained on these two classes, we do not report the classifications for these two supernova, however Table \ref{tab:SLSN} in Appendix \ref{app:res} details the tags associated with each. Whilst it is unclear from such a small sample whether any of the tags introduced in this work will be useful for identifying further SLSN, the lack of the hydrogen tag does agree with the definition of an SLSN-I.

%% file: Discussion.tex
\subsection{Importance of Tags}
From looking at Table \ref{tab:comp} and Table \ref{tab:comp2} of Appendix \ref{app:res} one can see that a certain classification can be due to the presence or absence of a particular tag or set of tags. If a spectrum has a high probability of a hydrogen tag then it will be classified as a Type II supernova almost regardless of the other tags that are present; as such the hydrogen tag can be considered to be highly dominant over all other tags. Similarly the presence of the silicon tag will mean that the classification is more than likely a Type Ia, though a relatively low probability can cause the confidence of such a classification to fall or result in an entirely different classification. Note that the presence of the other tags associated with a Type Ia (calcium H\&K, iron, sulphur) without the presence of the silicon tag is often not enough to select a Type Ia classification without other tags that are not typically associated with the Type Ia class (for example the helium emission tag with a high probability).

The tag for helium at 6678 \AA{} is also a very important tag, where even a very weak probability (below 10\%) can be enough to either reduce the confidence of the predicted class or changing the predicted class entirely. This is likely due to this tag rarely having a high probability (or indeed any probability) in any type of supernova that isn't a type IIb, and as such has a larger effect if it is present in any such supernova.

We do not find any common groupings of tags that are associated with the Type Ic class (as would be expected), instead spectra that lack spectral features or have combinations of tags that don't correspond to one class in particular will be put into this class. As such the Type Ic class can also be considered the class into which garbage spectra are currently placed into.

\subsection{Capability of \texttt{STag}}
\texttt{STag} shows reasonably strong agreement with \texttt{DASH} when classifying the OzDES spectra, with most of the differences being cause for further study to determine the true class. In such cases the list of tags, and their associated probabilities, allow a user to make a brief sanity check that the classification makes sense and provides the necessary justification should a classification be called into question. These tags are not only a means of justification but can also be used for purposes beyond just classification, such as selecting spectra that contains a particular feature of interest (provided it is included in the tags), which is not a guarantee with the current classification system.

Whilst both the ATel and \texttt{DASH} classifications do also give an estimate to the supernova's age, this capability is not present in the current version of \texttt{STag}. This functionality is planned to be added in future versions, though it is admittedly a shortcoming of the version presented in this paper. \texttt{STag} does have an advantage of scalability, in so far as tagging spectral features does not need to be limited to just supernova and could in practice be used on other non-supernova objects, which is also a feature planned to be implemented in the future.

%% file: Conclusion.tex
We present \texttt{STag}, a spectroscopic supernova classifier which uses logistic regression and multi-label classification to tag spectral features in supernova spectra and then uses these tag probabilities to a make classification by way of an artificial neural network. We use both template spectra collected from \texttt{DASH} and real spectra from OzDES in order to test the effectiveness of \texttt{STag} and find that it is capable of accurately classifying at minimum 95\% of the template spectra. We show good agreement with OzDES spectra previously classified by \texttt{DASH} (83\%) and identify possible previous misclassifications. We also perform new machine classifications using \texttt{DASH} in order to compare with \texttt{STag} and whilst we have only have a 39\% agreement, many of the \texttt{DASH} classifications are flagged as unreliable and as such we believe \texttt{STag} provides a better means of classification. 

Each spectral tag has an associated probability, which means that once the full process is complete a user has a comprehensive overview of a spectrum without the need for visual inspection. This can also be used to further refine and consolidate the way that supernovae are classified. \texttt{STag} is automated, requiring only an input spectrum and a redshift estimate, and is capable of classifying supernova at a competitive rate. With the expected large quantity of supernova discoveries to be made by surveys such as VRO, software like \texttt{STag} will be essential for the analysis of data from follow-up observations. To further improve this work we intend to increase the number of tags from the 10 currently used, which will not only improve the accuracy for the currently used classes but also allow us to include more sub-types. The use of spectral feature tags should be applicable to other transients or non-supernova objects (such as galaxies) and so there is room to expand the capability of \texttt{STag} to classify such objects.

%% file: res_table.tex
\startlongtable
\begin{deluxetable*}{ccccccp{0.23\linewidth}}
\tablecaption{Comparison of OzDEs ATel classifications, previous \texttt{DASH} classifications, and \texttt{STag} classifications.\label{tab:comp}}
\tablewidth{0pt}
\tablehead{
\colhead{OzDES} & \colhead{Observation Date} & \colhead{} & \colhead{ATel} & \colhead{\texttt{DASH}} & \colhead{\texttt{STag}} & \colhead{Feature}\\
\colhead{Name} & \colhead{(YY/MM/DD)} & \colhead{Redshift} & \colhead{Classification} & \colhead{Classification} & \colhead{Classification} & \colhead{Tag(s)}
}
\startdata
DES15S2ar & 15/09/19 & 0.247 & Ia? & Ia (0.977) & Ic (0.993) & Si (0.15), Ca (1.00), Fe (0.89), He-6678 (0.13)\\
\textquotedbl & 15/09/21 & \textquotedbl & \textquotedbl & \textquotedbl & Ic (0.986) & Ca (1.00), Fe (0.97)\\
DES15X2asq & 15/09/18 & 0.280 & Ia & Ia (0.655) & Ia (0.997) & Si (0.33), Ca (0.64), Fe (0.98), S (0.44), He-E (1.00), He-A (0.28)\\
\textquotedbl & 15/09/21 & \textquotedbl & \textquotedbl & \textquotedbl & Ia (1.000) & Si (1.00), Ca (0.43), Fe (0.99), S (0.87), He-E (0.97), He-C (0.96), He-A (0.41)\\
DES15C1atm & 15/09/18 & 0.207 & Ia & Ia (0.590) & Ia (0.993) & Si (0.80), Ca (1.00), Fe (1.00), He-C (0.79), H$\alpha$ (1.00)\\
DES15E2atw & 15/09/17 & 0.147 & Ia & Ia (0.911) & Ia (0.997) & Si (0.85), Ca (1.00), Fe (0.99)\\
\textquotedbl & 15/09/19 & \textquotedbl & \textquotedbl & \textquotedbl & Ia (0.995) & Si (0.90), Ca (1.00), Fe (0.94), H$\alpha$ (1.00)\\
\textquotedbl & 15/09/20 & \textquotedbl & \textquotedbl & \textquotedbl & Ia (0.888) & Si (0.99), Ca (1.00), Fe (0.93), H$\alpha$ (1.00), He-6678 (0.33)\\
DES15C2aty & 15/09/17 & 0.149 & Ia & Ia (0.405) & Ia (0.998) & Si (1.00), Ca (1.00), Fe (1.00)\\
\textquotedbl & 15/09/20 & \textquotedbl & \textquotedbl & \textquotedbl & Ia (0.998) & Si (0.99), Ca (1.00), Fe (1.00)\\
DES15X3auw & 15/09/17 & 0.151 & Ia & Ia (0.994) & Ia (0.999) & Si (1.00), Ca (1.00), Fe (1.00), H$\alpha$ (1.00)\\
\textquotedbl & 15/09/19 & \textquotedbl & \textquotedbl & \textquotedbl & Ia (0.999) & Si (1.00), Ca (1.00), Fe (1.00), H$\alpha$ (1.00)\\
DES15C3axd & 15/09/17 & 0.420 & Ia? & Ia (0.997) & Ia (1.000) & Si (1.00), Ca (0.98), S (0.71), He-E (1.00)\\
\textquotedbl & 15/09/20 & \textquotedbl & \textquotedbl & \textquotedbl & Ia (1.000) & Si (1.00), Ca (1.00), S (1.00), He-E (1.00), He-C (0.91), He-A (1.00)\\
DES15C3bj & 15/09/17 & 0.287 & II? & Ic (0.807) & II (1.000) & H (1.00), Ca (0.98), S (0.99), He-E (0.56), He-A (0.98), H$\alpha$ (1.00) \\
\textquotedbl & 15/09/20 & \textquotedbl & \textquotedbl & \textquotedbl & II (1.000) & H (1.00), Si (0.32), Ca (1.00), S (0.12), H$\alpha$ (1.00)\\
DES15X1bw & 15/09/20 & 0.130 & Ia & Ia (0.984) & Ia (0.983) & Si (1.00), Ca (0.36), Fe (0.99), He-C (0.11), He-A (1.00), H$\alpha$ (1.00)\\
\textquotedbl & 15/10/13 & \textquotedbl & \textquotedbl & \textquotedbl & Ia (0.927) & Si (1.00), Fe (0.99), He-C (0.50), He-A (1.00), H$\alpha$ (1.00)\\
\textquotedbl & 15/11/15 & \textquotedbl & \textquotedbl & \textquotedbl & Ia (0.927) & Si (1.00), Fe (0.98), He-C (0.97), He-A (1.00), H$\alpha$ (1.00)\\
DES15S1by & 15/09/19 & 0.129 & II & Ic (0.999)* & II (0.998) & H (1.00), S (0.93), He-E (0.99)\\
\textquotedbl & 15/09/20 & \textquotedbl & \textquotedbl & \textquotedbl & II (0.557) & H (1.00)\\
DES15S1cj & 15/09/19 & 0.166 & II? & II (0.832) & II (1.000) & H (1.00), Ca (0.84), Fe (0.46), He-A (1.00), H$\alpha$ (1.00)\\
\textquotedbl & 15/09/20 & \textquotedbl & \textquotedbl & \textquotedbl & II (0.995) & H (0.83), Ca (1.00), He-C (0.34), He-A (0.73), H$\alpha$ (1.00)\\
DES15E2cwm & 15/09/17 & 0.291 & Ia? & Ia (0.964) & Ia (0.427) & Si (0.40), Ca (1.00), He-A (1.00)\\
\textquotedbl & 15/09/20 & \textquotedbl & \textquotedbl & \textquotedbl & Ia (0.719) & Si (0.99), Fe (0.81)\\
DES15S2dye & 15/09/19 & 0.260 & Ia  & Ia (0.953) & Ia (1.000) & Si (1.00), Ca (1.00), Fe (0.99), S (0.99), He-E (1.00), He-C (1.00), He-A (1.00)\\
\textquotedbl & 15/09/21 & \textquotedbl & \textquotedbl & \textquotedbl & Ia (1.000) & Si (1.00), Ca (1.00), Fe (0.25), S (1.00), He-E (1.00), He-C (1.00), He-A (1.00)\\
DES15C2dyj & 15/09/17 & 0.395 & Ia & Ia (0.955) & Ia (1.000) & Si (0.95), Ca (1.00), S (1.00), He-E (1.00), He-A (1.00)\\
DES15X3dyu & 15/09/17 & 0.425 & Ia & Ia (0.938) & Ic (0.759) & H (0.78), Si (0.91), S (0.11), He-E (1.00)\\
\textquotedbl & 15/09/19 & \textquotedbl & \textquotedbl & \textquotedbl & Ia (1.000) & Si (0.97), Ca (1.00), S (0.88), He-E (1.00)\\
DES15C1eat & 15/09/18 & 0.450 & Ia? & Ia (0.612) & Ia (0.987) & Si (0.32), Ca (1.00), S (1.00), He-E (1.00)\\
DES15C2eaz & 15/09/17 & 0.062 & II & II (0.905) & II (1.000) & H (1.00), Si (0.81), He-E (0.99), He-C (0.19), H$\alpha$ (1.00), He-6678 (1.00)\\
DES15C1ebn & 15/09/18 & 0.410 & Ia? & Ic (0.904) & Ic (0.634) & Ca (0.31), S (1.00), He-E (1.00)\\
DES15X1ebs & 15/09/20 & 0.580 & Ia? & Ic (0.807) & Ic (0.999) & Ca (1.00)\\
DES15C3edd & 15/09/17 & 0.360 & Ia & Ia (0.707) & Ib (0.516) & Ca (1.00), S (1.00), He-E (1.00), He-A (0.90), He-6678 (0.11)\\
\textquotedbl & 15/09/20 & \textquotedbl & \textquotedbl & \textquotedbl & Ia (1.000) & Si (0.78), Ca (1.00), S (1.00), He-E (1.00), He-A (1.00)\\
DES15C3efn & 15/10/13 & 0.077 & Ia & Ia (0.964) & Ia (1.000) & Si (1.00), Ca (1.00), Fe (1.00), He-C (0.98)\\
DES15C3fx & 15/09/17 & 0.200 & Ia & Ia (0.999) & Ia (0.994) & Si (0.98), Ca (1.00), Fe (1.00), S (1.00), H$\alpha$ (1.00), He-6678 (0.99) \\
\textquotedbl & 15/09/20 & \textquotedbl & \textquotedbl & \textquotedbl & Ia (0.995) & Si (1.00), Ca (1.00), Fe (1.00), S (0.99), H$\alpha$ (1.00), He-6678 (0.99)\\
DES15X3hp & 15/09/17 & 0.236 & Ia & Ia (0.835) & Ia (0.938) & Si (1.00), Fe (0.98), He-C (0.76), He-A (1.00), H$\alpha$ (1.00)\\
\textquotedbl & 15/09/19 & \textquotedbl & \textquotedbl & \textquotedbl & Ia (0.981) & Si (1.00), Fe (0.98), He-C (0.62), He-A (0.75), H$\alpha$ (1.00)\\
DES15X3itc & 15/10/10 & 0.338 & Ia & Ia (1.000) & Ia (1.000) & Si (1.00), Ca (1.00), S (1.00), He-E (1.00), He-C (1.00), He-A (1.00)\\
\textquotedbl & 15/11/14 & \textquotedbl & \textquotedbl & \textquotedbl & Ia (1.000) & Si (1.00), Ca (1.00), S (1.00), He-E (1.00), He-C (1.00), He-A (1.00)\\
DES15X1ith & 15/10/13 & 0.160 & Ia & Ia (0.905) & Ia (0.995) & Si (1.00), Ca (1.00), Fe (1.00), He-6678 (0.16)\\
DES15E1iuh & 15/10/11 & 0.105 & II & II (0.918) & II (1.000) & H (1.00), Si (1.00), Ca (1.00), Fe (0.21), S (0.17), He-E (0.74), H$\alpha$ (1.00)\\
\textquotedbl & 15/10/13 & \textquotedbl & \textquotedbl & \textquotedbl & II (1.000) & H (1.00), Si (1.00), Ca (1.00), S (0.98), He-E (1.00), H$\alpha$ (1.00)\\
DES15X3iv & 15/10/10 & 0.018 & Ia & Ia (1.000) & Ia (1.000) & Si (1.00), Fe (1.00), He-C (1.00), He-A (1.00)\\
DES15X3kqv & 15/10/10 & 0.142 & Ia & Ia (0.996) & Ia (1.000) & Si (1.00), Ca (1.00), S (1.00), He-E (1.00), He-C (0.64), He-A (1.00)\\
DES15E1kwg & 15/10/11 & 0.105 & Ia & Ia (0.709) & Ia (1.000) & Si (1.00), Ca (1.00), S (1.00), He-E (1.00), He-C (0.72), He-A (1.00)\\
\textquotedbl & 15/10/13 & \textquotedbl & \textquotedbl & \textquotedbl & Ia (1.000) & Si (1.00), Ca (1.00), S (1.00), He-E (1.00), He-C (0.60), He-A (1.00)\\
DES15E1kvp & 15/10/11 & 0.442 & Ia & Ia (1.000) & Ia (1.000) & Si (0.88), Ca (1.00), S (1.00), He-E (1.00), He-C (1.00), He-A (1.00)\\
\textquotedbl & 15/10/13 & \textquotedbl & \textquotedbl & \textquotedbl & Ia (1.000) & Si (0.92), Ca (1.00), S (1.00), He-E (1.00), He-C (1.00), He-A (1.00)\\
DES15X3kxu & 15/10/10 & 0.345 & Ia & Ia (0.940) & Ia (1.000) & Si (0.97), Ca (1.00), S (1.00), He-E (1.00)\\
DES15C3lvt & 15/11/12 & 0.400 & Ia? & Ic (0.596)* & II (0.648) & S (1.00), He-E (0.95), He-C (0.60), He-A (0.91)\\
\textquotedbl & 15/11/13 & \textquotedbl & \textquotedbl & \textquotedbl & Ib (0.980) & Ca (0.95), S (0.30), He-E (0.66), He-C (0.58), He-A (1.00)\\
DES15E2nk & 15/09/17 & 0.308 & Ia & Ia (0.947) & Ia (1.000) & Si (0.99),  Fe (0.94), S (1.00), He-E (1.00), He-C (1.00), He-A (1.00)\\
\textquotedbl & 15/09/20 & \textquotedbl & \textquotedbl & \textquotedbl & Ia (1.000) & Si (0.99), Ca (0.19), Fe (0.92), S (1.00), He-E (1.00), He-C (1.00), He-A (1.00)\\
DES15S2og & 15/09/19 & 0.380 & Ia? & Ia (0.744) & Ia (0.999) & Si (0.79), Ca (1.00), Fe (0.78), S (0.45)\\
\textquotedbl & 15/09/21 & \textquotedbl & \textquotedbl & \textquotedbl & Ia (0.530) & Ca (1.00), Fe (0.15), S (1.00), He-E (0.15)\\
\enddata
\tablecomments{This table includes the OzDES designated name for the supernova, the date on which that particular spectrum was observed, the redshift as estimated by \texttt{\textsc{Marz}} \citep{Hinton2016}, the classifications from ATels, \texttt{DASH}, and \texttt{STag}, and a list of the all feature tags and their respective probabilities for each spectrum; with a requirement of the minimum probability to be 10\% in order to be included in the table. As such some spectra may have tags with probabilities below 10\% that are not included in this table. The tag names are based on the element responsible for that spectral feature; where the suffixes for helium are -E for emission, -C for P-Cygni, -A for absorption, and -6678 for the double peak at 6678 \AA{}.}
\end{deluxetable*}

%% file: new_res_table.tex
\startlongtable
\begin{deluxetable*}{ccccccp{0.23\linewidth}}
\tablecaption{Comparison of OzDEs ATel classifications, new \texttt{DASH} classifications, and \texttt{STag} classifications.\label{tab:comp2}}
\tablewidth{0pt}
\tablehead{
\colhead{OzDES} & \colhead{Observation Date} & \colhead{} & \colhead{ATel} & \colhead{\texttt{DASH}} & \colhead{\texttt{STag}} & \colhead{Feature}\\
\colhead{Name} & \colhead{(YY/MM/DD)} & \colhead{Redshift} & \colhead{Classification} & \colhead{Classification} & \colhead{Classification} & \colhead{Tag(s)}
}
\startdata
DES15S1asj & 15/09/19 & 0.140 & \nodata & IIb (0.843)* & II (0.977) & H (1.00), Ca (1.00), S (0.36), He-E (1.00), He-C (0.54), He-A (0.99), He-6678 (1.00)\\
\textquotedbl & 15/09/20 & \textquotedbl & \nodata & Ia (0.917)* & II (0.958) & H (1.00), Ca (0.83), S (0.10), He-E (1.00), He-C (0.74), He-A (0.99), He-6678 (1.00)\\
DES15X2asl & 15/09/18 & 0.324 & \nodata & Ia (0.995) & Ic (0.780) & H (1.00), Ca (0.13), S (1.00), He-E (0.60), He-6678 (0.55)\\
\textquotedbl & 15/09/21 & \textquotedbl & \nodata & Ia (0.767) & II (0.976) & H (1.00), Ca (1.00), S (1.00), He-E (1.00), He-6678 (0.19)\\
DES15X1aw & 15/09/20 & 0.329 & \nodata & II (0.785)* & II (0.980) & H (1.00), Si (0.59), Ca (0.99), He-C (1.00), He-A (1.00), H$\alpha$ (1.00)\\
DES15X1az & 15/09/20 & 0.151 & \nodata & Ia (0.677)* & IIb (0.859) & Ca (1.00), H$\alpha$ (1.00)\\
DES15E2bo & 15/09/17 & 0.233 & Ia & Ic (0.536)* & Ia (0.960) & Si (0.96), Ca (0.49), Fe (0.66), He-C (0.85), H$\alpha$ (1.00)\\
\textquotedbl & 15/09/20 & \textquotedbl & \textquotedbl & Ia (0.678) & Ia (1.000) & Si (0.96), Ca (1.00), Fe (1.00), He-C (0.84)\\
\textquotedbl & 15/10/11 & \textquotedbl & \textquotedbl & Ia (0.736) & Ia (1.000) & Si (1.00), Ca (1.00), Fe (1.00), He-C (0.97), He-A (0.49)\\
\textquotedbl & 15/10/13 & \textquotedbl & \textquotedbl & Ia (0.506) & Ia (1.000) & Si (0.99), Ca (1.00), Fe (1.00), He-C (0.98), He-A (0.54)\\
DES15S1ca & 15/09/19 & 0.180 & \nodata & Ia (0.995)* & II (0.906) & H (1.00), He-C (1.00), He-A (1.00), H$\alpha$ (1.00)\\
\textquotedbl & 15/09/20 & \textquotedbl & \nodata & Ia (0.914)* & II (1.000) & H (1.00), Si (0.70), He-C (0.81), He-A (1.00), H$\alpha$ (1.00)\\
DES15S1cs & 15/09/19 & 0.085 & \nodata & IIb (0.644)* & II (1.000) & H (1.00), Si (0.62), Ca (1.00), Fe (1.00), S (1.00), He-E (1.00), He-A (1.00), H$\alpha$ (1.00)\\
\textquotedbl & 15/09/20 & \textquotedbl & \nodata & Ic (0.918)* & II (1.000) & H (1.00), Ca (1.00), Fe (1.00), S (1.00), He-E (1.00), He-A (1.00), H$\alpha$ (1.00)\\
DES15C3cw & 15/09/17 & 0.207 & \nodata & Ic (0.994)* & Ia (0.998) & Si (0.96), Fe (0.20), He-E (1.00)\\
\textquotedbl & 15/09/20 & \textquotedbl & \nodata & Ic (0.994)* & Ia (1.000) & Si (0.99), Ca (0.87), He-E (1.00), He-C (0.19)\\
DES15C2dxc & 15/09/17 & 0.134 & \nodata & Ic (0.882)* & II (0.994) & H (0.99), Ca (1.00), He-E (1.00), H$\alpha$ (1.00), He-6678 (0.92)\\
DES15S1dyo & 15/09/19 & 0.487 & \nodata & II (0.950)* & II (0.937) & H (1.00), Si (0.31), Ca (1.00), S (0.99), He-E (1.00)\\
\textquotedbl & 15/09/20 & \textquotedbl & \nodata & Ia (0.906)* & Ia (0.568) & Si (0.38), S (0.99), He-E (1.00)\\
DES15E2dzb & 15/09/17 & 0.260 & Ia & Ia (0.933) & Ic (0.695) & Si (0.17), Ca (1.00), He-E (1.00)\\
\textquotedbl & 15/09/20 & \textquotedbl & \textquotedbl & Ia (0.948) & Ia (0.998) & Si (0.87), Ca (1.00), He-E (1.00)\\
\textquotedbl & 15/10/11 & \textquotedbl & \textquotedbl & Ia (0.905)* & Ia (1.000) & Si (0.99), Ca (1.00), S (0.53), He-E (1.00), H$\alpha$ (0.36)\\
\textquotedbl & 15/10/13 & \textquotedbl & \textquotedbl & Ia (0.977)* & Ia (0.995) & Si (0.99), Ca (0.96), Fe (0.26), S (0.90), He-E (1.00), He-C (0.62), H$\alpha$ (1.00)\\
\textquotedbl & 15/11/15 & \textquotedbl & \textquotedbl & Ia (0.996)* & Ia (0.502) & Si (0.67), Ca (1.00), S (0.88), He-E (1.00), He-C (0.38), H$\alpha$ (1.00), He-6678 (0.32)\\
DES15E2dzv & 15/09/17 & 0.337 & \nodata & II (0.862)* & Ia (0.632) & Si (0.30), S (0.94), He-E (1.00), He-C (0.36), He-A (1.00)\\
\textquotedbl & 15/09/20 & \textquotedbl & \nodata & IIb (0.999) & Ia (0.761) & Ca (1.00), S (0.98), He-E (1.00), He-C (0.99), He-A (1.00)\\
DES15E2eab & 15/09/17 & 0.635 & \nodata & Ic (0.507) & Ic (0.928) & Ca (1.00), S (0.93)\\
\textquotedbl & 15/09/20 & \textquotedbl & \nodata & Ic (1.000) & Ic (0.998) & Ca (0.99), S (0.13)\\
DES15S2eak & 15/09/19 & 0.430 & Ia & Ic (1.000) & Ic (0.707) & Si (0.83), Ca (1.00)\\
\textquotedbl & 15/09/21 & \textquotedbl & \textquotedbl & Ic (0.978)* & Ia (0.997) & Si (0.61), Ca (1.00), Fe (0.24), S (1.00), He-C (0.69), He-A (0.92)\\
DES15E1ece & 15/09/17 & 0.422 & Ia & II (0.997)* & Ic (0.848) & Si (0.50), He-C (0.89), He-A (0.97)\\
\textquotedbl & 15/09/19 & \textquotedbl & \textquotedbl & II (0.999)* & Ic (0.995) & Si (0.19), He-C (0.31), He-A (1.00)\\
DES15S1ecf & 15/09/19 & 0.052 & \nodata & Ic (0.998) & II (1.000) & H (1.00), Si (0.99), Fe (1.00), S (0.55), He-E (1.00), H$\alpha$ (1.00)\\
\textquotedbl & 15/09/20 & \textquotedbl & \nodata & Ic (0.772)* & II (1.000) & H (1.00), Si (0.94), Fe (1.00), He-E (1.00), H$\alpha$ (1.00)\\
\textquotedbl & 15/10/13 & \textquotedbl & \nodata & Ic (0.946) & II (0.999) & H (1.00), Si (0.16), Fe (1.00), He-E (1.00), H$\alpha$ (1.00)\\
DES15X2efu & 15/09/18 & 0.593 & \nodata & Ic (0.619) & Ic (0.999) & Fe (1.00), H$\alpha$ (1.00)\\
\textquotedbl & 15/09/21 & \textquotedbl & \nodata & Ic (0.726) & Ic (0.994) & Ca (1.00), Fe (1.00)\\
DES15X3fy & 15/09/17 & 0.435 & \nodata & II (0.993)* & II (0.993) & H (1.00), Ca (0.98), He-E (1.00), He-C (0.83), H$\alpha$ (1.00)\\
\textquotedbl & 15/09/19 & \textquotedbl & \nodata & II (0.950) & II (0.988) & H (1.00), Ca (0.51), He-E (1.00), H$\alpha$ (1.00)\\
DES15C1ge & 15/09/18 & 0.099 & \nodata & Ia (1.000) & IIb (0.933) & Si (0.91), He-C (0.73), H$\alpha$ (1.00)\\
DES15C1gm & 15/09/18 & 0.326 & \nodata & Ia (0.679) & Ic (0.817) & H (0.23), Ca (0.80), H$\alpha$ (1.00)\\
DES15C1gw & 15/09/18 & 0.256 & \nodata & II (0.999)* & IIb (0.999) & Ca (1.00), S (0.26), H$\alpha$ (1.00), He-6678 (1.00)\\
DES15C2ir & 15/09/17 & 0.059 & Ia & Ia (0.782) & Ia (0.955) & Si (1.00), Ca (1.00), Fe (1.00), He-C (1.00), He-A (0.98), H$\alpha$ (1.00)\\
DES15C3kc & 15/09/17 & 0.731 & \nodata & II (1.000)* & II (1.000) & H (1.00), Ca (1.00), He-E (0.23), H$\alpha$ (1.00)\\
\textquotedbl & 15/09/20 & \textquotedbl & \nodata & II (1.000) & II (1.000) & H (1.00), Ca (1.00), He-E (0.22), H$\alpha$ (1.00)\\
DES15C3kd & 15/09/17 & 0.296 & \nodata & Ia (0.829)* & II (0.987) & H (1.00), Ca (0.90), He-E (1.00), H$\alpha$ (1.00)\\
\textquotedbl & 15/09/20 & \textquotedbl & \nodata & Ia (0.490)* & II (1.000) & H (1.00), Ca (1.00), He-E (0.50), He-A (0.72), H$\alpha$ (1.00)\\
DES15C2kyh & 15/12/14 & 0.260 & Ia & Ia (0.999) & IIb (0.995) & Ca (1.00), Fe (1.00), He-C (1.00), He-A (0.19), H$\alpha$ (1.00), He-6678 (0.76)\\
DES15E1lbc & 15/12/13 & 0.127 & \nodata & Ia (1.000)* & Ia (0.508) & Si (0.50), Ca (1.00), He-E (1.00), H$\alpha$ (1.00)\\
DES15E1lei & 16/11/25 & 0.126 & \nodata & Ic (0.999) & IIb (0.958) & Si (0.86), Ca (1.00), H$\alpha$ (1.00)\\
\textquotedbl & 16/11/29 & \textquotedbl & \nodata & Ic (0.999) & IIb (0.983) & Si (0.61), Ca (1.00), H$\alpha$ (1.00)\\
DES15C2lp & 15/09/17 & 0.199 & \nodata & II (0.784)* & Ia (0.996) & Si (0.77), Ca (1.00), Fe (0.92), He-A (0.75)\\
DES15C3lrw & 15/12/12 & 0.128 & \nodata & IIb (1.000)* & IIb (1.000) & Si (0.40), Ca (1.00), H$\alpha$ (1.00), He-6678 (1.00)\\
DES15C3mgv & 15/12/12 & 0.306 & Ia & Ia (0.781) & Ia (0.997) & Si (1.00), Ca (1.00), Fe (0.91), He-C (0.22), H$\alpha$ (1.00)\\
DES15C3mud & 15/12/12 & 0.180 & \nodata & Ic (0.375)* & II (1.000) & H (1.00), He-E (0.99), He-A (1.00), H$\alpha$ (0.88)\\
DES15C1mvx & 15/12/13 & 0.534 & Ia & Ic (0.904) & IIb (0.992) & Ca (1.00), He-C (0.39), H$\alpha$ (1.00)\\
DES15C1myb & 15/12/13 & 0.419 & \nodata & Ia (1.000) & II (1.000) & H (0.98), Ca (1.00), S (1.00), He-E (1.00), He-C (1.00), He-A (1.00)\\
DES15C3naz & 15/12/12 & 0.331 & \nodata & Ia (1.000)* & Ia (1.000) & Si (1.00), Ca (1.00), He-E (1.00)\\
DES15E1nel & 15/12/13 & 0.599 & \nodata & Ic (0.424)* & Ic (0.999) & Ca (0.98)\\
DES15C2nfw & 15/12/14 & 0.337 & \nodata & Ic (0.905) & Ic (0.995) & He-E (0.99)\\
DES15C1nhv & 15/12/13 & 0.422 & Ia & Ia (0.962)* & II (0.962) & H (1.00), Si (0.56)\\
DES15C2npz & 16/02/06 & 0.121 & II & IIb (1.000)* & II (0.977) & H (1.00), Ca (1.00), He-E (0.99), He-C (0.26), H$\alpha$ (1.00), He-6678 (0.71)\\
DES15C3nqt & 16/02/06 & 0.138 & \nodata & II (0.757) & IIb (1.000) & Si (0.63), Ca (1.00), He-E (0.10), He-C (0.92), H$\alpha$ (1.00), He-6678 (0.68)\\
DES15C2odp & 16/02/06 & 0.336 & Ia & Ic (0.470)* & II (0.962) & H (1.00), Ca (0.97), He-E (1.00), He-6678 (0.42)\\
DES15C3odz & 16/02/06 & 0.509 & Ia & Ic (0.999)* & IIb (0.823) & Ca (1.00), H$\alpha$ (0.84)\\
DES15X1ol & 15/09/20 & 0.219 & \nodata & IIb (0.495)* & Ia (0.634) & Ca (1.00), Fe (1.00), He-C (0.74), He-A (0.97), H$\alpha$ (1.00)\\
DES15C3omh & 16/02/06 & 0.345 & Ia & Ia (0.998) & Ia (0.959) & Si (0.25), Ca (0.98), S (1.00), He-E (1.00), He-6678 (0.12)\\
DES15C2pgy & 16/02/06 & 0.535 & \nodata & II (1.000) & Ic (0.998) & Ca (0.96), Fe (0.54)\\
DES15E1q & 15/09/17 & 0.555 & \nodata & Ic (0.979)* & Ib (0.737) & Ca (1.00), Fe (1.00), He-C (0.63),  He-A (0.89)\\
\textquotedbl & 15/09/19 & \textquotedbl & \nodata & Ic (0.612)* & Ic (1.000) & No tags (above 10\%)\\
DES15C2sj & 15/09/17 & 0.070 & Ia? & Ic (0.848) & Ic (0.996) & Fe (1.00), S (1.00), He-6678 (1.00)\\
\enddata
\tablecomments{This table includes the OzDES designated name for the supernova, the date on which that particular spectrum was observed, the redshift as estimated by \texttt{\textsc{Marz}} \citep{Hinton2016}, the classifications from ATels, \texttt{DASH}, and \texttt{STag}, and a list of the all feature tags and their respective probabilities for each spectrum; with a requirement of the minimum probability to be 10\% in order to be included in the table. As such some spectra may have tags with probabilities below 10\% that are not included in this table. The tag names are based on the element responsible for that spectral feature; where the suffixes for helium are -E for emission, -C for P-Cygni, -A for absorption, and -6678 for the double peak at 6678 \AA{}.}
\end{deluxetable*}

%% file: SLSN_tab.tex
\begin{deluxetable*}{ccccp{0.23\linewidth}}
\tablecaption{Tags associated with SLSN-I supernovae as determined by \texttt{STag}.\label{tab:SLSN}}
\tablewidth{0pt}
\tablehead{
\colhead{OzDES} & \colhead{Observation Date} & \colhead{} & \colhead{ATel} & \colhead{Feature}\\
\colhead{Name} & \colhead{(YY/MM/DD)} & \colhead{Redshift} & \colhead{Classification} & \colhead{Tag(s)}
}
\startdata
DES15C3hav & 15/12/12 & 0.392 & SLSN-I & Ca (1.00), He-E (1.00), H$\alpha$ (1.00), He-6678 (0.10)\\
DES15S2nr & 15/09/19 & 0.220 & SLSN-I & S (1.00), He-E (1.00), He-A (0.10), He-6678 (1.00)\\
\textquotedbl & 15/09/21 & \textquotedbl & \textquotedbl & Si (0.86), Ca (1.00), Fe (0.80), S (1.00), He-E (1.00), He-C (0.21), He-A (1.00), H$\alpha$ (0.58) He-6678 (1.00)\\
\textquotedbl & 15/12/03 & \textquotedbl & \textquotedbl & Si (0.67), Ca (1.00), Fe (0.77), S (0.94), He-E (1.00), He-A (1.00), He-6678 (1.00)\\
\enddata
\tablecomments{This table includes the OzDES designated name for the supernova, the date on which that particular spectrum was observed, the redshift as estimated by \texttt{\textsc{Marz}} \citep{Hinton2016}, the classifications from ATels, and a list of the all feature tags and their respective probabilities for each spectrum; with a requirement of the minimum probability to be 10\% in order to be included in the table.}
\end{deluxetable*}